\documentclass{report}
\usepackage{graphicx}
\newcommand{\heart}{\ensuremath\heartsuit}
\graphicspath{ {images/} }
\usepackage[a4paper,width=150mm,top=25mm,bottom=25mm]{geometry}
\usepackage{amsmath,amssymb}
\usepackage{amsthm}

\RequirePackage[a-1b]{pdfx}
\RequirePackage[utf8]{inputenc}
\usepackage[utf8]{inputenc}

\usepackage[english]{babel}
\usepackage{cite}
\usepackage{parskip}

\usepackage{tabularx}
\usepackage{adjustbox}
\usepackage{array}
\usepackage[center]{caption}

\usepackage{fancyhdr}
\usepackage{url}
\usepackage{enumitem}
\RequirePackage{doi}
\usepackage{xcolor}
\usepackage{hyperref}
\hypersetup{
    colorlinks=false,  
    filebordercolor=red,
    linkbordercolor=blue, 
    urlbordercolor=green,
    citebordercolor=green,
    pdftitle={Monte-Carlo simulation of the Gaussian BFSS matrix model at large number of dimensions},
    pdfauthor={A.Haddad, B.Ydri},
}
\usepackage{caption}
\usepackage{subcaption}
\usepackage{epigraph}
\usepackage{multicol}
\usepackage{csquotes}
\usepackage{lmodern}
\date{\today}

\begin{document}

\begin{titlepage}
	\begin{center}
		\vspace*{1cm}
		\Huge
		\textbf{Monte-Carlo simulation of the Gaussian BFSS matrix model at large number of dimensions}
		
		\vspace{2cm}
		
		Defended on October 15, 2020
		
		\vspace{2cm}
		  
		\textit{Haddad Abdelhamid\\Supervised by Prof. Badis Ydri\\}
		\vspace{2cm}			
				Master Thesis on Theoretical physics
		
		\vspace{2cm}	
		\Large
		\begin{tabular}{ll}
		Bouchareb Adel & President \\
		Ramda Khaled & Examinator
		\end{tabular}
		
		\vfill
		
		Departement of physics\\
		Badji Mokhtar University\\
		Annaba, Algeria\\
		October 6, 2020.
		
	\end{center}
	\thispagestyle{empty}
\end{titlepage}
    \cleardoublepage
    \setcounter{page}{1}

\chapter*{\emph{Abstract}}
\addcontentsline{toc}{chapter}{Abstract} \markboth{Abstract}{}
\setcounter{page}{2}

\par In this thesis, we studied a Gaussian approximation to the bosonic part of the BFSS matrix model using Monte Carlo simulations based on Metropolis algorithm. We reproduces with great accuracy the stringy Hagedorn phase transition from a confinement (black string) phase to a deconfinement (black hole) phase. We used the Polyakov loop as an order parameter to investigate the large-N behaviour of this model at different temperatures, other observables such as internal energy and  extent of space were also computed. In last part, we present the matrix/geometry approach to a modified action where we captured only a remanent of the geometric Yang-Mills to a baby-fuzzy-sphere phase where the fuzzy sphere solution is only manifested as a three-cut configuration. The Yang-Mills phase retains most of its characteristics with two exceptions:  i) the uniform distribution inside a solid ball suffers a crossover at very small values of the gauge coupling constant to a Wigner’s semi-circle law, and ii) the uniform distribution at small T is non-existent.

\chapter*{\emph{Acknowledgements}}
\addcontentsline{toc}{chapter}{Acknowledgements} \markboth{Acknowledgements}{}

It is my dear pleasure and duty to dedicate these lines to all those who contributed, directly or indirectly to this work. You too who read this thesis could contribute to it in your own way, by indicating our errors and giving us your comments and I will be very honored by that (abdelhaddad.phy@gmail.com).

First of all, I would like to thank Professor Badis Ydri for having accepted to be my supervisor during this master thesis when I went to talk to him at the beginning of this year. Working with him has always been extremely challenging, as I always had the impression that his expectations were much higher than anything I could do. I apologize if sometimes I disappointed him by not being able to fulfill his expectations. It was also very enriching and allowed me to learn a lot of amazing things. Special thanks goes to My teacher, Dr. Adel Bouchareb which was of great importance when introduce me to the subject of the Ads/Cft correspondence last year and I really feel lucky to have been their student during this two years.

I also express my sincere gratitude to my Professors: R.Attalah, M.C.Talai, R.Chemam and M.Boulouednine for valuable advices, constant support and motivation throughout my two years at Annaba University.

I have also to thanks all the physics departement staff at Bouira University where I begun in a very great athmosphere with their company, whithout wich I will never be where I am today, they teach me a lot of what I need to think like a physicist.

Many thanks to Masanori Hanada at University of Surrey and S.Boukhalfa at Bouira University for many help at various stages of this thesis, to Govert Nijs at Utrecht University to have given me the opportunity to attend his doctoral defense and to read their dissertation and then open me the way to new horizons in link with the gauge/gravity duality.

I found great support from my fiends and classmates at Annaba as well as in Bouira University. In Particularly, A.Touati at Bouira University for multiple discussion and help, their comments on first versions of this thesis was very constructive. Without missing A.Goumidi, Z.Daoudi and many others for their hospitality at Constantine University.

I have also to thank "Alexandra Elbakyan" (the creator of Sci-Hub) without which many of the knowledge that I was in need to where simply unavailable...    

Finally, I want to thank my family. My mother and father may god keep them for me, as well as my brother Abdelraouf and wish him the best, also my maternel aunts for constant support and encouragement. 

I tried to be brief, I didn't mention them all. But I think the best way to thank them all is to succeed, first in what follow during this thesis, in PhD later and in my career even later... 
\begin{titlepage}

	\begin{center}
		\vspace*{1cm}
        \Huge
		
		\vspace{4cm}
		  
		\textit{To the men who taught me the meaning of the life, how to react and how to be.\\Died during the preparation of this thesis on 02/10/2020\\My father, my friend \heart\ .}
		\vspace{4cm}			
				
        \textit{To the bravest person I know,\\the most patient and beautiful\\may Allah bless him for me.\\A women like no other,\\My beloved mother.}
		
	\end{center}
	\thispagestyle{empty}
\end{titlepage}
    \cleardoublepage

\tableofcontents

\chapter*{\emph{Overview}}
\addcontentsline{toc}{chapter}{Overview} \markboth{OVERVIEW}{}

As mentionned briefly in the abstract above, we have tried to develop in this thesis the theory behind the Gaussian approximation of the bosonic BFSS Matrix model by our Monte carlo simulation of the model. Then after a brief introduction when we review together the "state of physics" in general way and try to give a guideline to how physicist have arrived to formulate this idea, we go in more futher details where we have devide the thesis into mainly three chapters as follow
\vspace{0.5cm}
\begin{itemize}

\item \textbf{The first chapter}, consist essentially on the needed theoretical background which we have constructed as
	\begin{itemize}
	\item \textbf{First section} which consists of an introductory chapter to the domain of string/M-theory, brane physics and supergravity, as well as the notion of Gauge/Graviy duality.
	\item \textbf{Second section} which is dedicated to the BFSS model itself when we try to present it as possible, their origin and some important remarqs are given first, then move to their large-d expansion which is identified as a Gaussian approximation of the bosonic model (which is the actual model we simulate), then we review their gravity dual briefly.
	\item \textbf{Third and laste section} Which aim to resume in a compact way what we gonna study numerically in next chapter, it may be useful for those who may like to pass the "beginners" details and get to the essence of the theory.
	\end{itemize}
	
\vspace{0.5cm}
\item \textbf{Second chapter} where we go more intensively in the computational part, we give the basics of our simulation first then gives and discuss our results and made some conclusions  
	\begin{itemize}
	\item \textbf{First section} present briefly the ideas behind the Monte Carlo simulation, how to think about it and how it works practically, then we explain how we have applied it to our Gaussian approximation of the bosonic BFSS model.
	\item \textbf{Second section} here we are in the heart of what we have done. We show, discuss and comments our simulation results and try to explain what occurs in the gravity dual theory each time is it possible to.
	\item \textbf{Third section} will resume the most important results and talk about we can do more in the near future.
	\end{itemize}

\vspace{0.5cm}
\item \textbf{Third chapter} treat the topic from the perspective of the matrix geometry approach. We begin by introducing some notion of non-commutative geometry and try to explain why fuzzy physics is important to us, then we move present our model briefly. We present the two phases of the model separetely adjointly with other observables then present a phase diagram wich compile them all and conclude the chapter by a section of conclusions.
\end{itemize}    
\chapter*{\emph{Introduction}}
\addcontentsline{toc}{chapter}{Introduction} \markboth{INTRODUCTION}{}

Friday, April 27, 1900 : \textit{"The beauty and clearness of the dynamical theory, which asserts heat and light to be modes of motion, is at present obscured by two clouds."} Was the words chosen by the British physicist Lord Kelvin when he gave a speech entitled \textit{"Nineteenth-Century Clouds over the Dynamical Theory of Heat and Light"} at the Royal Institution of Great Britain. This worlds give us the state of mind at this epoch where \textit{Heat} was understood by theormodynamics and statistical physics and the \textit{Light} by electromagntisme, it remains to perfect more and more precise measurement and to solve these \textit{"two clouds"} :
\vspace{0.5cm}
\begin{itemize}
\item The Michelson-Morley experiment which is the question of the luminous ether.
\item The ultraviolet catastrophe which is an obscur effect when studiying the black body radiation. 
\end{itemize}
\vspace{0.5cm}
These two clouds where as we will try to describe shortly here, the two pillar milestones of modern physics we know actually.

The first cloud, led us to formulate in 1905 the theory of \textit{Special Relativity} (SR) which first of all abandons the notion of absolute time, and \textit{General Relativity} (GR) in 1915 where events in the universe are not taking place on a fixed stage, but that spacetime itself is a dynamical quantity, and has to play an important part in the physics of the universe. John Archibald Wheeler have nicely resumed it when saying :
\vspace{0.5cm}
\begin{center}
\textit{"Spacetime tells matter how to move; matter tells spacetime how to curve."}
\end{center}
\vspace{0.5cm}
The second cloud, was explained by introducing in 1900 the idea of \textit{discrete "light quanta"} which invoke the concept of limitations on the allowed energy of emitted light and this was on the foundation of \textit{Quantum Mechanics} (QM) where one of the key ideas of this theory is the \textit{Wave-Particle duality} which claim that all things are both particles and waves at the same time and that nothing can be predicted or known with absolute certainty and the QM is a probabilistic theory.

The marriage of QM and SR leads to the framework of \textit{Quantum Field Theory} (QFT) which it proved to be an extremely successful language to describe all non-gravitational fundamental interactions in nature and described by quantum Yang-Mills field theories on a flat Minkowski four dimensional spacetime, the outcome of all that is the so-called \textit{Standart Model of particle physics} (SM) where matter are now knewed as fermions and forces between particles are mediated by another class of particles, so-called gauge bosons.

Two important notion we should talk about is that QFT are studied \textit{perturbatively} and it requires \textit{renormalisation} to be practically usable and predictive.

In view of the success of QFT in particle physics, it is tempting and desirable to believe that is it possible to describe all interactions by exchange of messenger bosons : including the gravitational force. But all the attempt to introduce a gauge boson mediating gravity by the so-called \textit{graviton} fails dramatically. Firstly, because this theory of \textit{Quantum Gravity} (QG) is pertubatively non-renormalizable, which mean that the computation of observables is plagued by an uncontrollable set of divergences which cannot be regulated and then the theory cannot predict anything! Secondly, as we have said earlier that in the theory of GR the spacetime is now dynamical and then it require a QFT wich are formulated in an independent fashion have to be rewritten in order to take the geometric nature of gravity into account.

More generally, the conceptual foundations of GR and QFT appear to be mutually incompatible: The former is a classical, strictly deterministic theory compatible with a dynamical spacetime, the latter theory incorporates quantum fields of intrinsically probabilistic nature but requires a fixed stage in order to define the fields. These conflicts suggest a more revolutionary approach to unification.

What we really need is a \textit{theory of everything} (TOE) : that includes all four fundamental forces and unifies "gravitic relativist" and "discrete quantum" theories. One of the most promising attempts to do this is \textit{string theory}, where particles are now considered as tiny one dimensional objects called strings. Just as a violin string can be made to vibrate in multiple modes. We have to admit that \textit{string theory} as such is not fully formulated yet, in particular not in a background independent or non-perturbative manner. 

Actually, there exist five perturbative string theories in 10-dimensions with a web of duality symmetries connecting them, and are now understood as different limits of a unified 11-dimensional theory, called \textit{M-theory}. Since the final form of a quantized M-theory is not yet known, most tests use the low-energy effective limit, which is 11-dimensional supergravity (a supersymmetric extension of GR). 

But not only that. In the late 90s another type of duality was found between string theory on a space called AdS space and conforal field theories living on the boundary of this space, the so-called \textit{AdS/CFT correspondence}. Generalized after that to \textit{Gauge/Gravity dualities}.

Modern attempts to formulate M-theory are typically based on matrix theory and the AdS/CFT correspondence.

In parallel, matrix model theories was developed historically to study \textit{Quantum Chromodynamics} (QCD) and have been used for two-dimensional QG. One important example of a matrix model is the BFSS matrix model proposed by Tom Banks, Willy Fischler, Stephen Shenker, and Leonard Susskind in 1997. This theory describes the behavior of a set of nine large matrices. In their original paper, these authors showed, among other things, that the low energy limit of this matrix model is described by 11-dimensional supergravity. Based on this fact, It aim to be a prototype of M-theory, and hence it conjecture that
\vspace{0.5cm}
\begin{center}
\textit{"M theory = M(atrix) theory."}
\end{center}

\chapter{\emph{Theory part}}

\epigraph{\textit{If superstring theory does turn out to be the TOE, historians of science will have a hard job explaining why it came into being.}}{Joel Shapiro.}

In this chapter, we will introduce step by step to how we came to the idea of matrix models in theoretical physics and how they are linked to M-theory. In first step, we will aboard the necessary needed background of \textit{string theory} to understand the general idea behind the \textit{BFSS matrix model} presented in next step. In the third and last step, we gives a brief conclusion for those who want to skip the details or already know it. 

\section{Prerequesties}

In this first section, the real goal is to review together the most important stages which are in concern with our subject\footnote{To be sincere, each of the steps mentioned here deserves a chapter all by itself, that's why we tried to be as clear and brief as possible, several references are given to accompany to fill in if needed.}. We first present a \textit{timeline} of the dates that we deemed the most important to construct a general picture of these ideas. After that, we try to decorticate them as we can in the view of this thesis. Introducing the notion of \textit{string and superstring theories} then moving to \textit{dualities} linking them. We give some important facts about \textit{M-theory and supergravity} which are in the heart of this thesis, then finish with talking about the \textit{Gauge/Gravity dualities}.
  
\subsection{Timeline}

As mentioned above, we'll present first in a form of a timeline the subjects linked with our thesis, see \cite{pais:1988,string0} for more historical details: 

\begin{description}
	\item{\textbf{1960s} :} String theory was constructed in an attempt to describe strong nuclear forces. Firstly introduced to explain the so-called \textit{Regge trajectories}\footnote{A linearly increasing spectra of hadronic mass versus angular momentum.} \cite{Regge1959IntroductionTC,PhysRevD.1.1182,Nambu:1997wf}.
	\item{\textbf{1970s} :} String theory was temporarily discarded because it contains a massless spin two particle (unwanted in a theory of hadrons) then pushed aside by QCD.
	\item{\textbf{1980s} :} The spin two excitation was identified with the graviton, and string theory turned out to be better suited for a more ambitious challenge: to serve as a theory of QG.
	\item{\textbf{1984-85} :} A first superstring revolution, string theory in 10-dimensional spacetime was shown to be free of quantum anomalies by M. Green and J. Schwarz \cite{Green:1984sg}. There was a series of discoveries \cite{Green:1984sg,gross1985heterotic,candelas1985vacuum} that convinced many theorists that superstring theory is a very promising approach to unification ans it was identified five different superstring theories.
	\item{\textbf{1994} :} Gerard 't Hooft speculate that the total number of degrees of freedom in a region of spacetime surrounding a black hole is proportional to the surface area of the horizon \cite{th93,th94}.
	\item{\textbf{1995} :} A second superstring revolution was triggered by the discovery of non-perturbative dualities due to E. Witten relating the five seemingly different theories and convince for the uniqueness of an underlying theory, the so-called M-theory \cite{witten1995}. In addition, an extra spatial dimension emerges in the spacetime of M-theory, and its low energy limit corresponds to the a unique supergravity theory in eleven dimensions \cite{string1}.
	\item{\textbf{1995} :} Susskind outline ’t Hooft’s idea and promote it to the so-called "holographic principle" \cite{susskind1994}.
	\item{\textbf{1997-98} :} Maldacena publishes an article that marks the start of the study of the AdS/CFT correspondence and thus provides a concrete realization of the cited holographic principle \cite{maldacena1997}. Generalized later to the so-called "gauge/gravity duality" \cite{aharony1999large}.
	\item{\textbf{1997} :} Tom Banks, Willy Fischler, Stephen Shenker, and Leonard Susskind showed that the low energy limit of their BFSS matrix model is described by eleven-dimensional supergravity. These calculations led them to propose that the BFSS matrix model is exactly equivalent to M-theory \cite{bfss}. 
\end{description}

\subsection{String and superstring theories}

The two strongest possibilities of theory of everythings (TOE) nowadays is the \textit{String theory} and the \textit{Loop Quantum Gravity} theories. We will not discuss the second one, our interest is in the string theory approach which suggest that all-known elementary particles are made up on buidling block: \textit{String} and the resonance of each string determine the properties of the particles (mass, charge, spin, ...) \cite{string3}.

To simplify their approach, physicist have considered at first stage the case of \textit{Bosonic string}, so we are in a world which contain only Bosons (forces), we will go back to how we included the fermions which constitute the matter of our universe later.

Before string theory, when we consider a point-like particle in space-time, their free trajectories will be a line, but for strings (one dimensional objects) the trajectories followed are a two dimensional surface in space time and form a \textit{World-sheet}. We also require the string to have a finite lenght, so it can either be an \textit{Open string} that can have gauge degrees of freedom at the edges, and hence it naturally describes the gauge fields \cite{string3}, or an \textit{Closed string} which describes among other things the so-called graviton.
\vspace{0.5cm}
\begin{center}
\captionsetup{type=figure}
\includegraphics[scale=0.99]{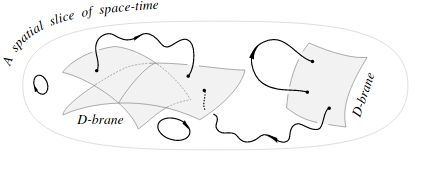}
\captionof{figure}{Different D-branes as boundary conditions for open strings in space-time along with closed string. Taken from \cite{string4}}
\end{center}

An important and purely mathematical notion when we deals with open string is the idea of \textit{D-brane} where D stay for \textit{Dirichlet} boundary conditions\footnote{They also exist exist a Neumann boundary condition, or either a mix between them: N-D or D-N, and these two type of boundaries appears when resolving the equations of motion of string theory.} and are a class of extended objects upon which open strings can ends in different manner showed in the figure above.

There is two crucial points appears in this theory of purely bosonic string, the first one is the dimention D of such string : for $D<26$ we found some negative probabilities $\psi^2<0$ and this problem dissapear at the critical dimension of $D=26$, the second one is that string theory was plagued by a tachyonic state in both the open and closed string sector violating causality and indicating an instability of the vacuum. The virtues of superstring theories established in the early 1980’s lie in the absence of tachyons \cite{Mmm}.

Now came the idea of \textit{Supersymmetry} (SYSY) which aim to transform a particle of a force (bosons) to matter (fermions) and vice-versa. SUSY on itself was born out of curiosity of finding a stabilisation mechanism to resolve the “hierarchy problem”\footnote{A hierarchy problem occurs when the fundamental value of some physical parameter, such as a coupling constant or a mass, in some Lagrangian is vastly different from its effective value (obtained after renormalisation prescription), which is the value that gets measured in an experiment.}, and as it turned, the possible resolution led to a spacetime
symmetry that exhanges boson for fermions

\begin{align}
\mathcal{Q}&|Boson>~~=~|Fermion>\nonumber\\
\mathcal{Q}&|Fermion>~=~|Boson>.
\end{align} 

Where $\mathcal{Q}$ is the SUSY fermionic generator i.e. they are spacetime spinors. As Q changes the spin of particles, it ultimately changes the properties of spacetime. Thus, SUSY is a spacetime symmetry that extends Poincare symmetry and its algebra.

The infinitesimal generator $\mathcal{Q}$ carry not only a spinor index $\alpha$, as $\mathcal{Q}^{\alpha}$. But it is possible to have more than one kind of SUSY transformation\footnote{Theories with more than one SUSY transformation are known as extended SUSY theories. The more extended SUSY is, the more it constrains physical observables and parameters of the theory.}. We distinguish them by adding an additional index $\textit{i}=1,2...\mathcal{N}$ as $\mathcal{Q}^{\alpha}_{\textit{i}}$.

Typically the number of copies of a supersymmetry is a power of 2 (1, 2, 4, 8...). In four dimensions, a spinor has four degrees of freedom and thus the minimal number of supersymmetry generators is four in four dimensions and having eight copies of supersymmetry means that there are 32 supersymmetry generators. 
The maximal number of supersymmetry generators possible is 32. Theories with more than 32 supersymmetry generators automatically have massless fields with spin greater than 2. It is not known how to make massless fields with spin greater than two interact, so the maximal number of supersymmetry generators considered is 32. This is due to the Weinberg–Witten theorem. This corresponds to an N = 8 supersymmetry theory. Theories with 32 supersymmetries automatically have a graviton. 

In reality when talking nowadays about string theory we mean \textit{super-string theory} and not the basic bosonic string theory and the critical dimention became $D=10$ (the dimention where we haven't negative propabilities anymore). 

Practically, there is five ways to produce the matter particle from the classical string theory where all have supersymmetry between forces and matter and cancel the tachyonic state !
 
\begin{itemize}
\item Type I : With both closed and open strings, group symmetry is $SO(32)$.
\item Type IIA : Closed and open strings bound to D-branes, massless fermions spin both ways (nonchiral). 
\item Type IIB : Closed and open strings bound to D-branes, massless fermions only spin one way (chiral).
\item Heterotic $E_{8}×E_{8}$ : With closed string only, which is a hybrid of a superstring and a bosonic string (Heterotic), symmetry group is $E_{8}×E_{8}$.
\item Heterotic $SO(32)$ : Closed strings only, heterotic, group symmetry is $SO(32)$.
\end{itemize}

We will see in next section how these five "different" superstring theories are related to each other and get us to came up with the idea of M-theory.

\subsection{Dualities, M-theory, Supergravity}

So to resume last section, there is five main (super-)string theories living in 10-dimentional space-time. But in reality, the five superstring theories are connected by a web of dualities and then are regarded as different limits of a single theory, the so-called \textit{M-theory}. This remains a conjecture and we haven't, in properly saying the formulation of such theory : the goal of the BFSS matrix model is to try to emphasize it, and in this view-point are called the M(atrix) theory, we will go back later to this point. 

\begin{center}
\captionsetup{type=figure}
\includegraphics[scale=0.39]{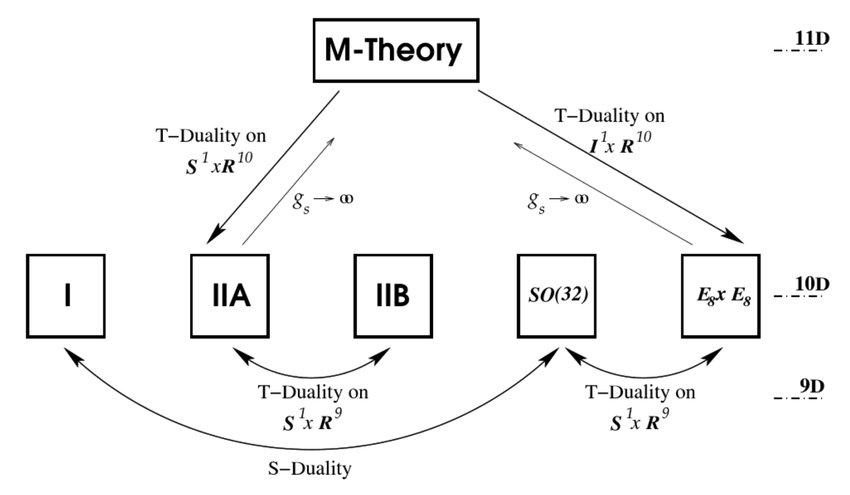}
\captionof{figure}{The web of S- and T-dualities connecting the five 10-dimensional superstring theories and 11-dimensional M-theory. Taken from \cite{M-theory2}.}\label{duality}
\end{center}

The first which came up with this idea of M-theory is Witten in the spring 1995 in a conference at the University of Southern California \cite{witten1995}. Witten's based his conjecture part on the intuition from the fact that the five versions of superstring theory although appeared, at first, to be very different, but related in and intricate and nontrivial ways. Physicists found that apparently distinct theories could be unified by mathematical transformations called \textit{S-duality} and \textit{T-duality}, represented in figure \ref{duality}, see \cite{duality} and \cite{M-theory2} for technical details.

\begin{itemize}
\item \textit{The S-duality}, where the S are for "Strong" : stay that a theory A with strong coupling between strings is equivalent to a theory B with weak coupling.

\item \textit{The T-duality}, where the T are for "Target" : stay that a theory A with large compact dimension is equivalent to a theory B with small dimension.

\item There is also an combined duality between them named the \textit{U-duality} which for example, can be regarded as a transformation that exchanges a large geometry of one theory with the strong coupling of another theory.
\end{itemize}

Another pillar which will be of great importance in next chapter is the \textit{supergravity theories} (SUGRA). In essence, supergravity combines the principles of SUSY and of GR into one field theory. Skipping a lot of details that can be found in \cite{string1,string2}, we will resume their importance into two point: 

\begin{itemize}
\item There is multiple supergravity theories at differents dimensions, but there is one supergravity in 11-dimensional space. 
\item Due to last string dualities, the conjectured 11-dimensional M-theory is required to have 11-dimensional supergravity as a "low energy limit".
\end{itemize}

So there is different "configuration" to construct a SUGRA theory. The configurations of interest to us are those arising as the classical solution to 10D and 11D SUGRA, which are known to respectively provide the low energy effective field theory of superstring theories and M-theory.

\begin{center}
\captionsetup{type=figure}
\includegraphics[scale=0.37]{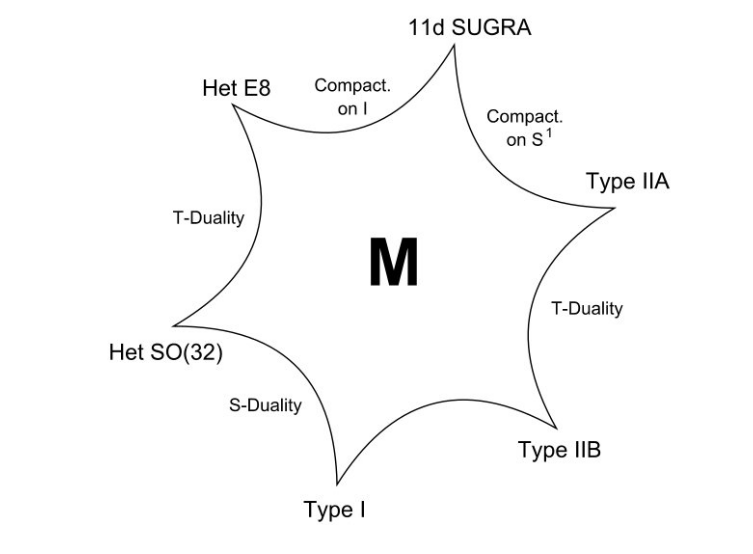}
\captionof{figure}{Map of Various String theories related by string dualities, the 11d SUGRA limit of M-theory is also shown - Wiki.}\label{mm}
\end{center}

Note however the fact that different string theory gives different SUGRA. In example, the low-energy effective theory of type IIA superstring theory is  
type IIA supergravity, etc. Also, the low-energy effective theory of M-theory is 11d SUGRA. As a final remark, as various string theories are related by dualties: SUGRA theories are also related to each other with dualities. This is summarized in figure \ref{mm} ,see \cite{SUGRA} that resume this very well.  

All these dualities we talk about before are between differents theories, but the culminent point of all these ideas of dualities is a duality between open and closed strings itself. This will be presented in next section and are mostly knowed as the \textit{holographic duality}.

\subsection{Gauge/Gravity dualities}

The various Gauge/Gravity theory dualities claim that certain quantum theories of gravity in ($d+1$)-dimensional backgrounds are equivalent, or dual, to certain quantum field theories in d dimensions. In many cases this quantum field theory is a gauge theory which led to the name "gauge/gravity duality".  For instance, the holographic principle\footnote{Which was inspired originally by the Bekenstein-Hawking formula for the black hole entropy \cite{anagnostopoulos2007}, see \cite{luminet2016holographic,bousso2002holographic} for details.}.

\begin{center}
\captionsetup{type=figure}
\includegraphics[scale=0.34]{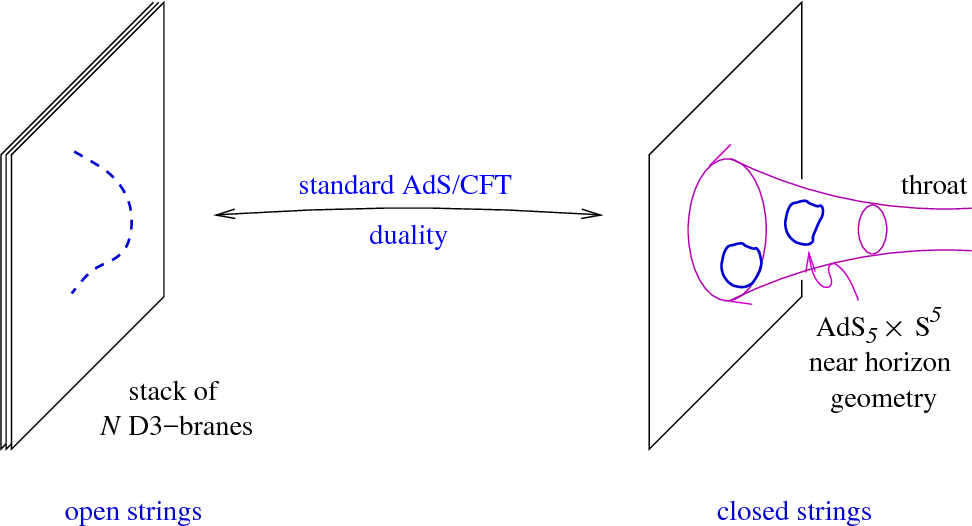}
\captionof{figure}{Standard AdS/CFT: The left figure shows the description of a stack of D3-branes in terms of open strings (Yang-Mills description), the right figure in terms of closed strings (supergravity description). Taken from \cite{ads-cft}.}
\end{center}

A concrete example of a Gauge/Gravity duality is the Ads/Cft correspondance where

\vspace{0.5cm}

\begin{center}

\textit{Four-dimensional $\mathcal{N} = 4$ supersymmetric SU(N) gauge theory is equivalent to IIB string theory with $AdS_{5} \times S^{5}$ boundary conditions.}
 
\end{center}

\vspace{0.5cm}

See \cite{maldacena1997} for details. We can resume it as an equivalence between the gravity theory living in the \textit{bulk} side expressed by graviton (closed string), and the gauge theory living in the \textit{boundaries} (open strings attached to D-brane) that express the three other forces.

This duality also open up a window onto the strong coupling dynamics of gauge theories. In recent times they have been applied to a variety of physical systems that range from high-energy phenomenology to condensed matter physics \cite{conden,hartnoll,ammon}.

keep in mind that a reasonable overview of all past and current developments would require a textbook on its own, so we have omit a great number of details. Our aim is to present enough material to convey a general picture of this subject needed for this thesis, we will end this section by a picture that came in my mind during this writing and I really like to share it here with you. 

Analogies between physical theories and other subjects like Art, economy or life situations in general was always of great interest for me. Maybe some of you have experienced the fact that when reading a roman, essays or more generally\footnote{I've talk here about reading espacially because it's what happen with me the most, but it can also be true with watching TV or others.} a text which are not really from their domain of work to distress themselves that our mind try always to search a comparison between them (maybe unconsciously buts it's here). Thinking how I can explain the idea of gauge/gravity duality which are in the heart of this thesis make me remember a text that I've read in relation with psychology from an austrian psychotherapist named Alfred Adler:
\vspace{0.5cm}
\begin{displayquote}
<<We should not be astonished if in the cases where we see an inferiority [feeling] complex we find a superiority complex more or less hidden. On the other hand, if we inquire into a superiority complex and study its con­tinuity, we can always find a moie or less hidden inferiority [feeling] complex.>>\cite{adler}
\end{displayquote}
\vspace{0.5cm}
We'll not go more deeply to what is a "superiority/inferiority complex", but we'll take the idea that in certain case, we can found fragment of a theory in another one which really didn't suggest it at the beginning...

\section{The BFSS model}

In this second section, we will review the D0-brane matrix model first, then gives some primary results obtained by leading papers. Some remarqs about large N limit and 't Hooft coupling are given. We then move to present the harmonic oscillator version from the initial model by a large d expansion which is what studied numerically in our thesis. We will end by talking about the gravity dual of the model.

\subsection{the Model}

As seen in last section, the 11-dimensional M-theory aim to combine the five "different" 10-dimensional superstring theories \cite{witten1995}. which are in fact related by duality transformations, and merge into the idea of a single fundamental theory: M-theory \cite{makeenkothree}. As the five superstring theories have as a low-energy limit a SUGRA theory. Thus M-theory itself should have an 11-dimensional supergravity solution as a low-energy limit. T.Banks, W.Fischler, S.Shenker, and L.Susskind (BFSS) showed in \cite{bfss} that the low-energy limit of their matrix model is described by 11-dimensional supergravity. And it is the key point! They conjecture that the BFSS model in the limit of infinite matrices can be used as a prototype for a correct formulation of M-theory and thus became, 
\vspace{0.5cm}
\begin{center}
						M theory = M(atrix) theory
\end{center}
\vspace{0.5cm}
The easiest way to obtain the BFSS model, also known as the D0-brane matrix model\footnote{The reason is that the model is supposed to describe a collection of D0-branes. We can think the d0-branes as a "single points".} is via dimensional reduction of 10-dimensional SYM theory down to one dimension.
\vspace{0.5cm}
\begin{center}
              9 space + 1 time $\longrightarrow$ 0 space + 1 time.
\end{center}
\vspace{0.5cm}
We will skip the entire procedure given in the appendix of \cite{tanwar} and gives the action explicitly by\footnote{Here we have performed a wick's rotation in prevision of what follow, the reason of this step will be clear in next chapter. The original action is obtained simply by putting a plus sign before the commutator term!}

\begin{eqnarray} 
S=\frac{1}{g^2}\int_0^{\beta}dt{\rm Tr}\bigg[\frac{1}{2}(D_t\Phi_i)^2-\frac{1}{4}[\Phi_i,\Phi_j]^2+Fermions\bigg].\label{BFSS} 
\end{eqnarray}

Where $Tr$ is trace over matrices, $D_{t} \equiv \partial_{t}-i[A(t), . ]$ represents the covariant derivative. The one-dimensional fields $\phi_{i}$(t) (i = 1, 2...9) which is an $N \times N$ Hermitian matrices and the field $A(t)$ can be regarded respectively as the gauge field and the nine adjoint scalars of the model. The only free parameter in this model is the Hawking temperature $T = 1/\beta$.

The full supersymmetric model \eqref{BFSS} has been intensively studied in the literature, see pioneer paper \cite{catterall2008,kawahara2007,anagnostopoulos2007}, for more precision see \cite{kadoh2015,filev2015} and more recently \cite{hanada-test}. The main goal of these studies has been to compare the low-temperature regime to the holographically dual black hole geometry obtained by SUGRA theory\footnote{Presented shortly at the end of this chapter.}. Their results are in reasonable agreement for large temperature but disagree at lower temperature ($T < 1$), they observed small deviation in comparison with SUGRA predictions, see figure \ref{test}.
\vspace{0.5cm}
\begin{center}
\captionsetup{type=figure}
\includegraphics[scale=0.4]{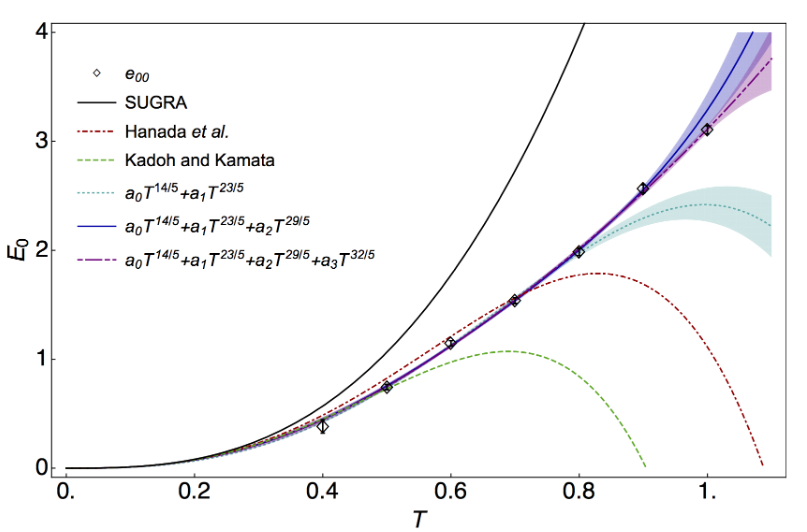}
\captionof{figure}{Precision lattice test of the gauge/gravity based on the numerical computation of internal energy of the black hole directly from the gauge theory where they reproduce the coefficient of the supergravity result $\frac{E}{N^{2}}=7.41T^{14/5}$, including some stringy corrections. The SUGRA result is shown in black. The result from \cite{kadoh2015} and \cite{hanada2008higher} results are respectively given by the green dashed and red dot-dashed lines. Taken from \cite{hanada-test}}\label{test}
\end{center}

As a result, we can say that the BFSS model reproduce well known result of 11D SUGRA for finite N and temperature. Then constitute a non-trivial check of Gauge/gravity duality, and maybe in future can predict the finite quantum stringy correction based on simulation! 

In what follow, we will be interested in the \textit{Energy} $\frac{E}{N^{2}}$ too, but also in such obervables as the \textit{Polyakov line} $<|P|>$ and the \textit{Extent of space} $R^{2}$ which act as order parameters for the model and that will be defined later. the eigenvalue distribution of the Holonomy $A(t)$ and the field $\phi_{i}$ will also be of great interest.
					
One of the questions that our understanding completely fall down, is the discovery of the fact that black holes radiate particles and eventually evaporate: the well-known \textit{information loss paradox} \cite{hawking1975}. This paradox caused a long and serious debate since it claims that the fundamental laws of quantum mechanics may be violated. A possible cure appears if we conjecture that the holographic description of a quantum black hole based on the gauge/gravity duality is correct, the information is not lost and quantum mechanics remains valid. 

Here \cite{small1,small2} they test this gauge/gravity duality on a computer at the level of quantum gravity. The black hole mass obtained by Monte Carlo simulation of the dual gauge theory reproduces precisely the quantum gravity effects in an evaporating black hole. This result opens up totally new perspectives towards quantum gravity since one can simulate quantum black holes through dual gauge theories.

\subsection{The 't Hooft and large-N limits}

This point is pretty delicate but important for us, their explicit explanation is a bit far from what we aim to present, But we will resume it as we understand and you can refer to \cite{thooft2002large,coleman1985,manohar1998large} and \cite{witten1979large} for more explicit details.

The situation is such that for classes of gauge theories such as SYM or matrix models, whose gauge groups may be $N \times N$ square matrices (like $SU(N)$ for QCD). The properties of the theory is given when taking the \textit{“large number of colours-limit”} := $N\to\infty$ that scales as 1/N and allow a perturbation series around \textit{the large-N limit}, the  so-called the \textit{1/N expansion}.

In QCD, this \textit{1/N expansion} serves to provide a computational tool for describing confined hadron states which are not seen by ordinary perturbation theory in the gauge theory coupling constant.

Remind that at early stages, string theory was developped in aim to describe strong force, laterly pushed aside by QCD. But the "war" don't end at this point. ’t Hooft have made an important step towards out understandig of both theories, he noticed that SU(N) gauge theories with a large N simplify\footnote{The diagrams that dominate in the large N limit are those with the minimum number of quark loops.}, and he also conjectured that in this limit QCD is described by a string theory. 

This limit provides a hint to and closely relate QFT to string theory, and then plays an essential role for the AdS/CFT correspondence which provides a map between them and we may look it as,
\vspace{0.5cm}
\begin{center}
\textit{Large N Gauge Theories as String Theories.}
\end{center}
\vspace{0.5cm}
An extreme case of this is the large N-limit of the BFSS matrix model where all spatial dependence of fields in the higher dimensional spacetime is supposedly encoded in the quantum mechanics of $N\times N$ matrices as $N\to\infty$.

In preview mentioned procedure\footnote{You really need to return to references cited above to understand this point more clearly.}, it will turn out to be extremely useful to define the ’t Hooft coupling by:  

\begin{eqnarray}
\lambda := g_{YM}^{2}N_{c}
\end{eqnarray}

Where $g_{YM}$ is the coupling constant for Yang-Mills theory and with and $N_{c}$ colours. 

The ’t Hooft large-N limit corresponds to sending $N$ to $\infty$  with the ’t Hooft coupling constant $\lambda \equiv g^{2}N$ fixed, this procedure is sometimes called the \textit{“double scaling limit”}. Since the coupling constant $g$ can be absorbed by rescaling the matrices and time appropriately, we can set $\lambda$ to unity without loss of generality. This implies that we replace the prefactor $\frac{1}{g^{2}}$ in the action by N in what follows. Adjoining to that, we will also perform a $1/d$ expansion.

\subsection{1/d expansion : The matrix harmonic oscillator}

We will pick the M-(atrix) theory action for quenched fermions of the full model \eqref{BFSS},

\begin{eqnarray} 
S=\frac{1}{g^2}\int_0^{\beta}dt{\rm Tr}\bigg[\frac{1}{2}(D_t\Phi_i)^2-\frac{1}{4}[\Phi_i,\Phi_j]^2\bigg].\label{bosonicBFSS} 
\end{eqnarray}

But our scope is more modest and we will only concern ourselves with the model after performing a large number of dimensions $d\longrightarrow\infty$ expansion (assuming $d=9$ is quite large)\footnote{A 1/d expansion is considered in lattice models originally in \cite{Drouffe:1979dh,DROUFFE19831}. See the related treatment in \cite{kabat2000black,kabat2001black}.}. The model \eqref{bosonicBFSS} became equivalent to a matrix harmonic oscillator problem given by the following simple matrix scalar field theory, see \cite{filev2015} and \cite{gauss1} for details,

\begin{eqnarray}
S[\Phi]=\frac{1}{g^2} \int_0^\beta dt Tr \bigg[ \frac{1}{2} (D_t\Phi_i)^2 + \frac{1}{2} m^2 (\Phi_i)^2 \bigg]~,~m=d^{1/3}. \label{gaussian}  
\end{eqnarray}

It has been argued in \cite{filev2015} that the dynamics of the bosonic BFSS model \eqref{bosonicBFSS} is fully dominated by the large d behavior encoded in the above quadratic action. And a part of our work have been to check this in Monte Carlo simulations. 

\subsection{The gravity dual}

This section is slightly different from what we presented above, and are putted here for desire of completeness. You can consult \cite{grav0,grav1,grav2} for more details.

\begin{center}
\captionsetup{type=figure}
\includegraphics[scale=0.925]{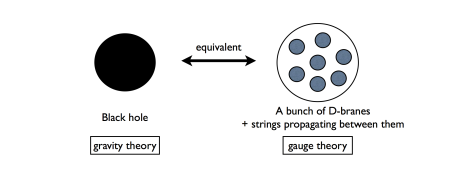}
\captionof{figure}{Black holes in superstring theory are conjectured to be described by the dual gauge theory. Taken from \cite{small2}.}
\end{center}
\vspace{0.6cm}
We've talked a bit about the gauge side above. Let us start with an analysis of the gravity side. We will discuss it in three stages.

\begin{enumerate}

\item The black hole, which is made of N D-particles in superstring theory (In our model, it is a bunch of D0-branes), is described by a curved ten-dimensional space-time, which can be obtained as a solution to the classical equation of motion (or the “Einstein equation”) for supergravity \cite{small2}. The energy at this stage was explicitely computed in \cite{grav2} as, 

\begin{eqnarray}
\frac{1}{N^2}E_{Classical} = 7.41T^{2.8}.
\end{eqnarray}

\item We can now consider quantum corrections to this classical geometry. Since superstring theory is defined perturbatively, one can calculate the leading quantum corrections to the geometry\footnote{Which correspond to the $1/N^{2}$ corrections.} by explicitly solving the equations of motion for the near horizon geometry as it was done in \cite{hyakutake2013quantum},

\begin{eqnarray}
\frac{1}{N^2}E_{Gravity} = 7.41T^{2.8}-5.77T^{0.4} \frac{1}{N^{2}}+\mathcal{O}(\frac{1}{N^{4}}).
\end{eqnarray}

\item There is also the so-called $\alpha '$ which represent the effects due to the lenghts and oscillations of strings,

\begin{eqnarray}
\boxed{\frac{1}{N^{2}}E_{Gravity}^{Total} = (7.41T^{2.8}+aT^{4.6}+\cdots)+(-5.77T^{0.4}+bT^{8.2}+\cdots)\frac{1}{N^{2}}+\mathcal{O}(\frac{1}{N^{4}}).}
\end{eqnarray}

The other corrections a,b,etc are actually unknown, and we expect the simulations to provide a tool to compute them numerically. see figure \ref{test} again for to convince yourself.

\end{enumerate}
\vspace{0.5cm}
To conclude, we say "simply" that the BFSS matrix model are belived to describe the IIA string around black 0-brane (near horizon) in the gravity dual theory. If correct, It should reproduce thermodynamics of this black 0-brane. The simulations can be performed in two different manner,
\vspace{0.2cm}
\begin{itemize}

\item Simulations at large N so that string loop corrections can be neglected on the gravity side. The results \cite{small1,hanada-test,kadoh2015,anagnostopoulos2007} for the internal energy exhibit the temperature dependence consistent with the prediction including the $\alpha '$ corrections.

\item We can also perform simulations at small N but at lower temperature so that the $\alpha '$ corrections can be neglected on the gravity side. The results \cite{small1,small2} are consistent with the prediction including the leading string loop correction, which suggests that the conjecture holds even at finite N.

\end{itemize}

\section{Summary}

\begin{itemize}

\item There are 5 manifestation of superstring theories, all consistent in 10 spacetime dimensions and related by various string dualities. String duality has also led to the prediction of an 11D theory called M-theory. The classical limit of superstring theories as well as M-theory are respectively given by 10D and 11D supergravity theories.

\item Gauge/gravity duality based on the holographic principle which is a conjecture between gravitational and non-gravitational theories provide (if correct!) an equivalence such that, 
\vspace{0.4cm}
\begin{center}
\boxed{\text{Closed string theory (gravity) = Open string theory (gauge theory).}}
\end{center}
\vspace{0.5cm}
\item If this is true, the (lattice) gauge field theory may provide a nonperturbative
formulation of the closed string theory, which is a theory of quantum gravity! Since the gauge/gravity duality is confirmed including quantum gravity effects, we can study various issues involving quantum gravity by using Monte Carlo simulation of the dual gauge theory.

\item The BFSS model reproduce well known result of 11D SUGRA for finite N and temperature, and constitute a non-trivial check of Gauge/gravity duality.

\item Our present model is a d gauged matrix harmonic oscillators, viz.

\begin{eqnarray} 
\boxed{S[\Phi]=\frac{1}{g^2}\int_0^{\beta} dt Tr\bigg[\frac{1}{2}(D_t\Phi_i)^2+\frac{1}{2}m^2(\Phi_i)^2\bigg], m=d^{1/3}.}\nonumber  
\end{eqnarray}

\item At finite temperature the gravity dual has the geometry solution of ten  dimensional type IIA supergavity closely related to the near horizon geometry of a charged black hole in ten dimensions. Their energy is expected to be reproduced by the gauge/gravity duality from the dual gauge theory.

\end{itemize}

\chapter{\emph{Simulation I: Gauge/Gravity perspective}}

\epigraph{\textit{Physics tells us observations can't be predicted absolutely. Rather, there's a range of possible observations each with a different probability.}}{Robert Lanza.}

Remember that the topic of this thesis is the numerical simulation of our Gaussian approximation of the bosonic BFSS matrix model. We've seen the theory, now let's turn to the simulation part. The goal of this chapter is to present the basic skills we need to start carrying out Monte Carlo studies, as well as to way we have apply it of our topic and finally present and discuss the results.

\section{Monte Carlo}

Let's point first the numerical method used in our simulation, the so-called \textit{Monte-carlo}\footnote{Which refers to the Monte Carlo Casino in Monaco.} alorithm. The subject of Monte Carlo and in general the subject of \textit{lattice field theory}\footnote{Lattice field theory is the study of lattice models of QFT on a spacetime that has been discretized onto a lattice. Although most of them are not exactly solvable, one hopes that, by performing simulations on larger and larger lattices, while making the lattice spacing smaller and smaller, one will be able to recover the behaviour of the continuum theory.} is "vast" and it is a domain by itself on ArXiv platform. We will try to be brief and consice as possible, we will follow \cite{ydricomputational,hanadamarkov,anosh}.

Markov Chain Monte Carlo (MCMC) are a subset of computational algorithms that use the process of repeated random sampling (random number generation) to make numerical estimations of unknown parameters and to solve deterministic problems. But before explaining MCMC, we should explain how do we need it first!

\subsection{Intergal Approximation}

Two classes of statistical problems are most commonly addressed within this framework: integration and optimization: it is the (approximate) calculation of integrals using collections of random samples that people usually think of when they refer to the Monte Carlo Method.

Imagine we have a function of p variables $f(x_{1},x_{2}...x_{p})$ and we want to integrate it. First we can say that we can approximate this integral by means of the rectangular approximation as in figure \ref{integr}, the integral is simply equal to the area (or volume) of bins (in each direction).

\begin{center}
\captionsetup{type=figure}
\includegraphics[scale=0.65]{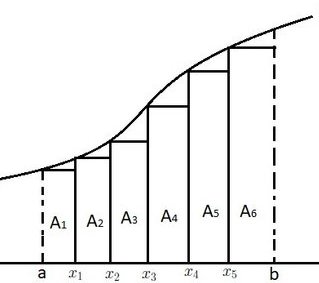}
\captionof{figure}{Approximation of an intagral of a function of p-variable by a summation over areas of bins (rectangular approximation).}\label{integr}
\end{center}

We can say that if we make the bin finer and finer: the result converges to the right value. It is no more than a definition of what an integral is.

But imagine we have $10$ variables, and construct $100$ bin for each direction, than we would have $10^{20}$ bins, and we have to sum over it to solve this integral ! The situation is getting worse if we go to more variables or bins.\footnote{To give you an idea, one of the fastest computer in the world \textit{"Livermore Lab Sequoia"} would solve this probleme, when for $10^{20}$ operation, we need 5000 secondes (still acceptable), for $10^{30}$ around 634 years. (many thanks to M.Hanada talk given at ICTS Bangalore in 2018 under the title "A Numerical Approach to Holography").}

MCMC solve this this problem by using the notion of \textit{important simpling}. Imagine we are trying to do the path integral of some physical system: 

\begin{eqnarray} 
Z = \int {\cal D}x e^{-S[x]}.\label{function} 
\end{eqnarray}

Where here the last function $f(x_{1},x_{2},...,x_{p}) \equiv e^{-S[x]}$, you can imagine this action $S[x]$ as the potential illustrated as the altitude of the Grand Canyon in figure \ref{cany}.

\begin{center}
\captionsetup{type=figure}
\includegraphics[scale=0.45]{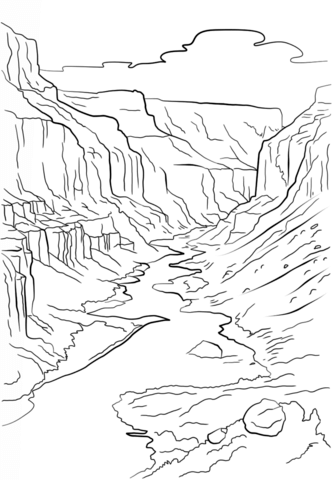}
\captionof{figure}{The Grand Canyon Illustration.}\label{cany}
\end{center}

And we have then a huge phase space, intuitively we can say that not all region have the same importance. We have some region which does not contribute too much to the integral. We should not wast out "limited" ressources to cumpute such region, which correspond to unimportant configurations of the physical system. But we should only pick important bins which contibute significantly to the integral: This is the idea of \textit{Important sampling}.

\subsection{Important sampling}

But the question that arises now, is how the MCMC actually do it ? the answer is based using random numbers to do such integral. 
 
On the rest, consider a field theory on an Euclidean space-time\footnote{Euclidean field theory be a unification of relativistic field theory with statistical mechanics/thermodynamics, then called \textit{thermal QFT} or \textit{quantum statistical field theory} or similar. Such theories are obtained by the so-called \textit{wick's rotation}: which involves analytic continuation of n-point functions to complex valued time coordinates by the substitutions $t\longrightarrow -i\tau$ which gives $iS\longrightarrow -S_E$ (it is the reason of the definiton \ref{function}).} with the action $S[\phi]$ then we generate field configurations with probabilities $\propto \exp(-S)$ (probability wheight) and "important" samples are created more then others, then for an observales $\mathcal{O}$

\begin{eqnarray}
< \mathcal{O} > = \frac{\int {\cal D}\phi \mathcal{O} [\phi] e^{-S[\phi]}}{\int {\cal D}\phi e^{-S[\phi]}} \approx \frac{1}{n} \sum_{i=1}^{n} \mathcal{O} [\phi_{i}]
\end{eqnarray}

Are no more than an arithmetic average when the number of sample $n$ are very large. Such a set of configurations can be generated as long as the assupmtion that this probabilty is real and positive $\exp(-S[\phi])>0$\footnote{When is becomes negative or imaginary we can't use the MCMC methods, the so-called \textit{"Sign problem"} that typically occurs in theories of fermions when the fermion chemical potential $\mu$ is nonzero.}.

The idea is to generate a chain of field configurations $C$ with a transition probability $P[C \to C']$ with the condition that the series

\begin{eqnarray}
C_0 \to C_1 \to C_2 \to ... \nonumber
\end{eqnarray}

does not depend on the starting point (configuration) $C_0$ and the transition from the $k_{th}$ to $K_{th+1}$ does not depend on the previous history $0_{th}$ to $k_{th-1}$ configurations, it is the reason why is called it a \textit{Markov Chain}.

Different MCMC can differ from our choice of how we generate new configurations $P[C \to C']$ and how we create this famous chain. For our probleme which contain boson only, the most simple and commonly used technique for obtaining updates is used, the so-called \textit{Metropolis algorithm}. For other system which include fermions, we need methods like \textit{Hybrid Monte Carlo} (HMC) and \textit{Rational Hybrid Monte Carlo} (RHMC), see references cited above.

\subsection{Metropolis Algorithm}

The Metropolis algorithm can be summarized as follows:

\begin{enumerate}
\item Choose an initial field configuration (initialisation).
\item Choose a matrix element from a lattice point n and make a random change.
\item Compute $\Delta S=S_{Trial}-S_{Original}$. This is the change in the energy of the system due to last update.
\item Check if $\Delta S \leq 0$. In this case the trial is accepted.
\item Check if $\Delta S > 0$. In this case compute the ratio of probabilities $\omega = e^{-\Delta S}$.
\item Choose a uniform random number $r$ in the inetrval $[0,1]$.
\item Verify if $r \leq \omega$.  In this case the trial is accepted, otherwise it is rejected.
\item Repeat steps 2) through 7) until all elements of the system are tested. This sweep counts as one unit of Monte Carlo time.
\item Repeat  setps  2)  through  8)  a  sufficient  number  of  times  until  thermalization, i.e. equilibrium is reached and we should not obtain configurations with small probabilities.
\item Compute the physical quantities of interest in $n$ thermalized configurations. This can be done periodically in order to reduce correlation between the data points.
\item Compute averages.
\end{enumerate}

the steps (2) through (7) corresponds to a transition probability between the microstates ${S_{i}}$ and ${S_{j}}$ given by

\begin{eqnarray} 
W(i \longrightarrow j) = {\rm min}(1,e^{-\Delta S}), \Delta S = S_{j}-S_{i}.  
\end{eqnarray}

Since only the ratio of probabilities $\omega = e{-\Delta S}$ is needed to generate a sequence of states which are distributed according to the Boltzmann distribution.

\subsection{Coding steps}

After having introduced the techniques and algorithms needed, we can now discuss how to organize the actual simulation. A Monte Carlo simulation consists of several basic steps summarized as follows,

\subsubsection{Lattice regularization}

Field theoretic calculations are made more explicit and more rigorous by working on an Euclidean lattice spacetime. In fact, the Euclidean lattice provides a concrete non-perturbative definition of the theory. Thus, we will employ lattice regularization in which

\begin{align}
&\phi^i(x)~=~\phi_n^i(n).\\
&\int dt~=~a \sum_{n=1}^{\Lambda}.\\
&t~=~(n-1)a~,~(n = 1,...\Lambda)~\text{with}~\beta~=~\Lambda a.
\end{align} 

Where the index "n" in $\phi_n^i(n)$ correspont to the lattice point, "a" represent the lattice spacing and $\Lambda$ the total number of lattice points. This means in particular that the lattice can be throught actually as a circle, this is illustrated in figure \ref{circle}.

\begin{center}
\captionsetup{type=figure}
\includegraphics[scale=0.65]{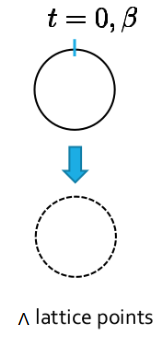}
\captionof{figure}{From continuum to Lattice discretisation.}\label{circle}
\end{center}

Where the lattice field $\phi_n^i(n)$ occupy now the lattice points, and the nine dimension $d=9$ means that we have in each site $9$ matrices $(i=1...9)$, each of them have $N \times N$ elements.

The lattice treatement of the gauge-field $A(t)$ and the covariant derivative will be given in the gauge-fixing part below.

\subsubsection{Gauge fixing}

Let's remind the previous definition of the covatiant derivative $D_{t} \equiv \partial_{t} - i[A(t), . ]$ given in \eqref{BFSS}. Where the interactions of the scalar fields $\Phi_i(t)$ with a one-dimensional $U(N)$ gauge field $A(t)$ is implemented in the usual way through the minimal coupling

\begin{eqnarray} 
\partial_t\Phi_i\longrightarrow D_t\Phi_i=\partial_t\Phi_i-i[A,\Phi_i]. 
\end{eqnarray}

Taking our model given in \eqref{gaussian} which is our gaussian approximation of the bosonic part of the BFSS, is invariant under the $U(N)$ gauge transformations

\begin{eqnarray} 
\Phi_i\longrightarrow U\Phi_iU^{\dagger}~,~A\longrightarrow UAU^{\dagger}-\frac{1}{i}U\partial_tU^{\dagger}. 
\end{eqnarray}  

We will now gauge-fix this symmetry non-perturbatively \cite{filev2015}. Taking into account what we did previously in this section, the gauge field A becomes the link variable (also called the transporter fields) defined as

\begin{eqnarray} 
U_{n,n+1}={\cal P}\exp(-i\int_{na}^{(n+1)a}dt A(t)).\label{link} 
\end{eqnarray}

where ${\cal P}$ denotes a path ordered product and $U_{n+1,n} = U_{n,n+1}^{\dagger}$. To make the covariant derivative gauge covariant, we have to transport back the field at $t_{n+1}$ to $t_{n}$. For the discrete version of the covariant derivative, we obtain

\begin{eqnarray} 
D_t\Phi_i = \frac{U_{n,n+1}\Phi_i(n+1)U_{n+1,n}-\Phi_i(n)}{a}.
\end{eqnarray}

The discrete gaussian action becomes

\begin{eqnarray} 
S=\frac{1}{g^2}\sum_{n=1}^{\Lambda}{ Tr}\bigg[\frac{1}{a}\Phi_i^2(n)-\frac{1}{a}U_{n,n+1}\Phi_i(n+1)U_{n+1,n}\Phi_i(n)+\frac{1}{2}m^2a\Phi_i(n)^2\bigg]. 
\end{eqnarray}

Alternatively\footnote{This procedure is not so simple, it invokes the fact that we have a local $U(N)$ symmetry at each lattice site and that we can rotate the link variables safely, the entire procedure is well explained in \cite{filev2015,ydriblog}.}, this action can also be rewritten in terms of $D={\rm diag}(\exp(i\theta_1),...,\exp(i\theta_N))$\footnote{This because we have choose a static diagonale gauge $A(t)=-(\theta_1,\theta_2,...,\theta_N)/\beta$ and where the $\theta_a$ are defined as the holonomy angles.} as

\begin{eqnarray} 
S&=&\frac{1}{g^2}\sum_{n=1}^{\Lambda}{ Tr}\bigg[\frac{1}{a}\Phi_i^{2}(n)+\frac{1}{2}a^2m^2\Phi_i^{2}(n)\bigg]\nonumber\\ 
&-&\frac{1}{g^2a}\sum_{n=1}^{\Lambda}{Tr}\Phi_i(n)D_{\Lambda}\Phi_i(n+1)D_{\Lambda}^{\dagger}.\label{final}
\end{eqnarray}

Where $D_{\Lambda}=D^{1/\Lambda}$.

Finally, we reduce the measure over the transporter fields as follows

\begin{eqnarray} 
  \prod_{n=1}^{\Lambda}{\cal D}U_{n,n+1}&=&\prod_{n=2}^{\Lambda}{\cal D}U_{n,n+1}{\cal D}U_{1,2}\nonumber\\ 
  &\sim&\prod_{a=1}^Nd\theta_a.\prod_{a>b}|e^{i\theta_a}-e^{i\theta_b}|^2\nonumber\\ 
  &\sim&\prod_{a=1}^Nd\theta_a.\prod_{a>b}\sin^2\frac{\theta_a-\theta_b}{2}\nonumber\\ 
  &\sim &\prod_{a=1}^Nd\theta_a.\exp(-S_{\rm FP}). 
\end{eqnarray}

Where $S_{\rm FP}$ is the so-called \textit{Faddeev-Popov} gauge-fixing action, and it is given explicitly by

\begin{eqnarray} 
\boxed{S_{\rm FP}=-\frac{1}{2}\sum_{a\ne b}\ln\sin^2\frac{\theta_a-\theta_b}{2}.}\label{faddeev-popov}
\end{eqnarray}

Then the Gauge-fixing procedure is to define properly the covariant derivative with the gauge field as in \eqref{final}, and add to him the Faddeev-Popov term \eqref{faddeev-popov}. In summary, we are interested in the total action

\begin{eqnarray} 
S_{\rm total}&=&N\sum_{n=1}^{\Lambda}{
 Tr}\bigg[\frac{1}{a}{\Phi}_i^{ 
2}(n)+\frac{1}{2}am^2{\Phi}_i^{ 2}(n)\bigg]\nonumber\\ 
&-&\frac{N}{a}\sum_{n=1}^{\Lambda}{Tr}{\Phi}_i^{}(n)D_{\Lambda}{\Phi}_i^{}(n+1)D_{\Lambda}^{\dagger}-\frac{1}{2}\sum_{a\ne b}\ln\sin^2\frac{\theta_a-\theta_b}{2}.\label{last}
\end{eqnarray}

\subsubsection{Boundary condition}

Since a numerical simulation works on a finite lattice, boundary conditions have
to be implemented. This matrix harmonic oscillator is a thermal field theory in one dimension and thus $t$ must be  an imaginary time, such that $\tilde{t}=-it$ is the real time (See the foot-note about wick's rotation). The fields are periodic with period $\beta=1/T$ where $T$ is the Hawking temperature as 

\begin{eqnarray} 
\Phi_i(t+\beta)=\Phi_i(\beta). 
\end{eqnarray}

Since the system are studied in a lattice, the periodicity condition becomes

\begin{eqnarray} 
\boxed{\Phi_i(n+\Lambda)=\Phi_i(n).}
\end{eqnarray}

\subsubsection{Evaluation of the observables}

Before we discuss the results, we present the most important observables. 

\begin{itemize}
\item The Polyakov line $\langle |P|\rangle$ is the order parameter of the Hagedorn transition in string theory and the deconfinment transition in gauge theory which is associated with the spontaneous breakdown of the $U(1)$ symmetry $A(t)\longrightarrow A(t)+C.{\bf 1}$, see \cite{kawahara2007phase} for details. The Polyakov line is defined in terms of the holonomy matrix $U$ given in \eqref{link} by the relation

\begin{eqnarray} 
P=\frac{1}{N}Tr U~,~U={\cal P}\exp(-i\int_0^{\beta} dt A(t)).  
\end{eqnarray}

Hence

\begin{eqnarray} 
\boxed{P=\frac{1}{N}\sum_ae^{i\theta_a}.}  
\end{eqnarray}

\item Another important observable is the radius (or extent of space or more precisely the extent of the eigenvalue distribution) defined by

\begin{eqnarray} 
\boxed{R^2=\frac{a}{\Lambda N^2}\langle {\rm radius}\rangle~,~{\rm radius}=\frac{N}{a}\sum_{n=1}^{\Lambda}{Tr}{\Phi}_i^{2}(n).}
\end{eqnarray}

\item The energy of the bosonic truncation of BFSS matrix model \eqref{bosonicBFSS} is given normally by 

\begin{eqnarray} 
\frac{E}{N^2}=\frac{3T}{N^2}\langle
 {\rm commu}\rangle~,~{\rm 
commu}=-\frac{1}{4g^2}\int_{0}^{\beta}dt{Tr}[{\Phi}_i^{},{\Phi}_j^{}]^2. 
\end{eqnarray}

But because of our large d expansion, the model becomes \eqref{gaussian} and the actual energy is given effectively by the extent of space as

\begin{eqnarray} 
\boxed{\frac{E}{N^2}=\frac{a^2Tm^2}{N^{2}}\langle{\rm radius}\rangle=m^2R^2.}\label{energ}
\end{eqnarray} 

\item As a complement, we also measure the eigenvalues distribution of the holonomy matrix $U$.

\end{itemize}

\subsubsection{Initialization}

Any field configuration can be chosen as initial configuration. After reaching the
equilibrium, configurations will be distributed according to the Boltzmann distribution. For our simulation, we have the scalar field $\phi_i(n)$ and the holonomy angles $\theta_a$.

We have choose a \textit{Cold start}\footnote{A Hot start correspond to take the matrices elements randomly.} for the field $\phi_i(n)$, viz.
 
\begin{eqnarray}
\boxed{\phi_i(n)=0.}
\end{eqnarray}

In order for numerical efficiency, we have takes the static diagonal gauge,

\begin{eqnarray}
A(t)=-(\theta_1,\theta_2,...,\theta_N)/\beta
\end{eqnarray}

in such a way that

\begin{eqnarray}
\pi < \theta_a \leq -\pi
\end{eqnarray}

This comes from the holonomy angles variation of the action (see appendix), and that the Polyakov line phases are constrained in this interval. 

\subsubsection{Numerical errors}

We use the Jackknife Method to evaluate the errors. This part of the code can be found in \cite{ydricomputational}. In the Jackknife method, we first divide the configurations to bins with width $\omega$, the first bin is $x^1 \to x^\omega$, the second $x^\omega \to x^{2\omega}$, etc. Then we define the average of an observable $f(x)$ with $k_{th}$ bin removed 

\begin{eqnarray}
\bar{f}^{(k, \omega )} \equiv (\text{the value calculated after removing $k_{th}$ bin}).
\end{eqnarray}

The average value is defined by

\begin{eqnarray}
\bar{f} \equiv \frac{1}{n} \sum_{k} \bar{f}^{(k,\omega)}. 
\end{eqnarray}

The Jackknife error is defined by

\begin{eqnarray}
\boxed{\Delta_{\omega} \equiv \sqrt{\frac{n-1}{n} \sum_{k}(\bar{f}^{(k,\omega)}-\bar{f})^{2}}.}
\end{eqnarray}

\subsubsection{Final remarks}

We write the code according along the above lines using Fortran77\footnote{The potential superior features which may be found in C are peripheral to our purposes here. No more that our "simple" Laptop was used and was sufficient to us.}. We run simulations for $N = 10, 12, 16$. We use typicaly $2^{14}$ thermalization steps and $2^{14}$ measurements steps. 

Depending on the updating algorithm, consecutively generated configurations will be more or less correlated. Thus, in principle, several sweeps of order $2^{6}$ are performed between diffrent measurements.

The code was written from scratch. Still it is extremely rare to use anything more than $+,-,\times,\div$ and $\sin,\cos,\exp,\log$, “if” and loop. Some linear algebra routines from LAPACK was also needed to compute the eigenvalues of the bosonic matrices $\Phi_i(n)$, but you can find it in \cite{ydricomputational}.

For the random number, we use the Ran2 generator. Subroutines can be found in \cite{ydricomputational}.

\section{Phase Structure}

In this second section, we will present some of our Monte Carlo results, they are also summarized mostly in \cite{ydri2020}, see \cite{filev2015} for complement. 

For what follow, we will fix $d=9$ as required by the theoretical model, and the lattice spacing to $a=0.05$. The free parameter of the model is the temperature $T$, the number of lattice point will be defined numerically by setting $\Lambda = \frac{1}{T.a}$ and will vary as the temperature vary!

The phase diagram of this model was determined numerically by means of the Monte Carlo method in \cite{filev2015} to be consisting of two phase transitions and three stable phases, this will be cheked below.

\subsection{Polyakov line and Holonomy angles}

First, we can check the behaviour of the Polyakov line $\langle |P|\rangle$ which plays the rôle of an order parameter for the confining-deconfining phase transition given in figure \ref{polyakov}\footnote{At high temperatures the bosonic part of the BFSS quantum mechanics reduces to the bosonic part of the IKKT model (reduced one dimension further down to D(-1)-branes in type IIB string theory), the leading behavior of the various observables of interest at high temperatures can be obtained in terms of the corresponding expectation values in the IKKT model. This behavior is essentially used to calibrate Monte Carlo simulations at high temperatures.\cite{ydri2017review}}
\vspace{0.5cm}
\begin{center}
\captionsetup{type=figure}
\includegraphics[scale=0.5]{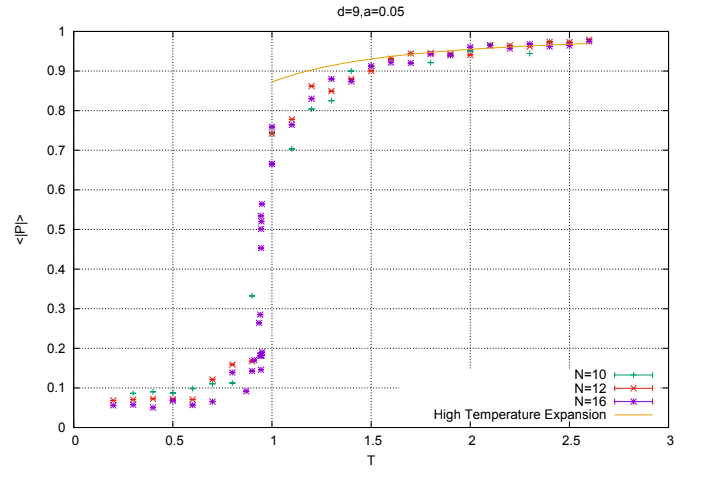}
\captionof{figure}{A plot of the Polyakov line for $N=10, 12, 16$ at large number of dimensions $(d=9)$ for lattice $a=0.05$ as a function of the temperature. The high temparature expansion is obtained from \cite{kawahara2007}.}\label{polyakov}
\end{center}
\vspace{0.5cm}
At first approximation, we can see that near temperature $T \approx 0.9$ there is a first order phase transition (like a discontinous "jump") in the confuguration of the $\theta_{a}$'s.  but a more detailed analysis of the temperature range close to the transition reveal that there are in fact two transitions. From low to high T as

\begin{enumerate}

\item A second order phase transition from a uniform to a non-uniform distribution of the $\theta_{a}$'s angles defined as confinement/deconfinement transition (at a Hagedorn temperature $T_{H}=T_{c2})$. Closely followed by a:

\item A third order phase transition which translate into the emergence of a gap in the eigenvalue distribution of the $\theta_{a}$'s angles. 

\end{enumerate}

The first transition (the Hagedorn/Deconfinement Phase Transition) will be discussed in a separate section below. 

For the second transition, and more generally in matrix field theories, this type of transitions in which the eigenvalues distribution develops a gap are knowed more generally as a Gross-Witten-Wadia phase transitions (GWW), you may consult the appendix the end of this thesis for details.

But for now, let's made theses phases more clearer as follow in plots of figure \ref{theta} numerically,

\begin{center}
\captionsetup{type=figure}
\includegraphics[scale=0.33]{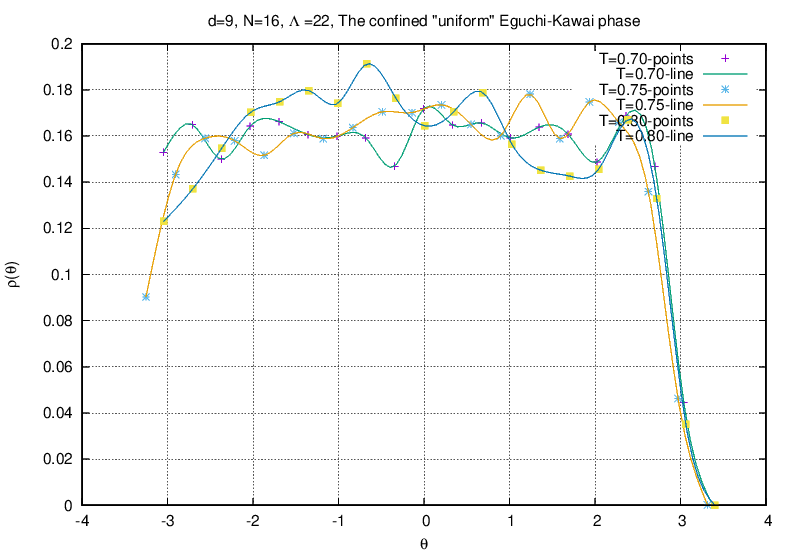}
\includegraphics[scale=0.33]{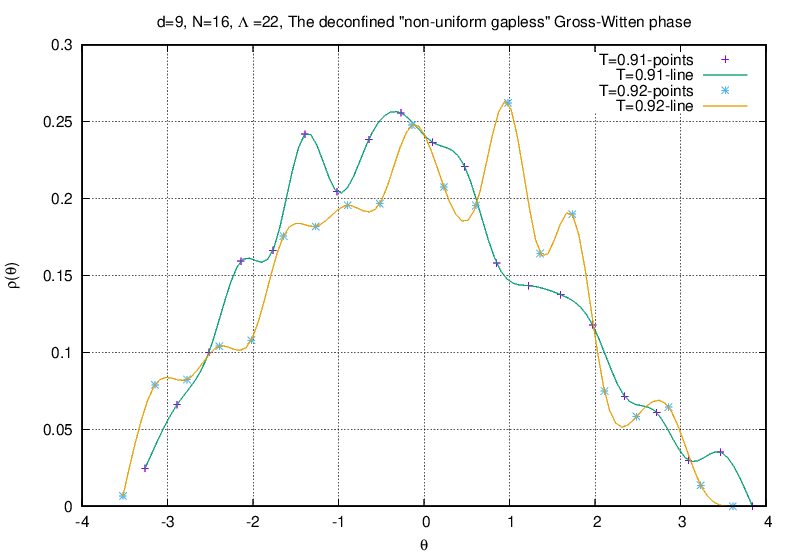}
\includegraphics[scale=0.33]{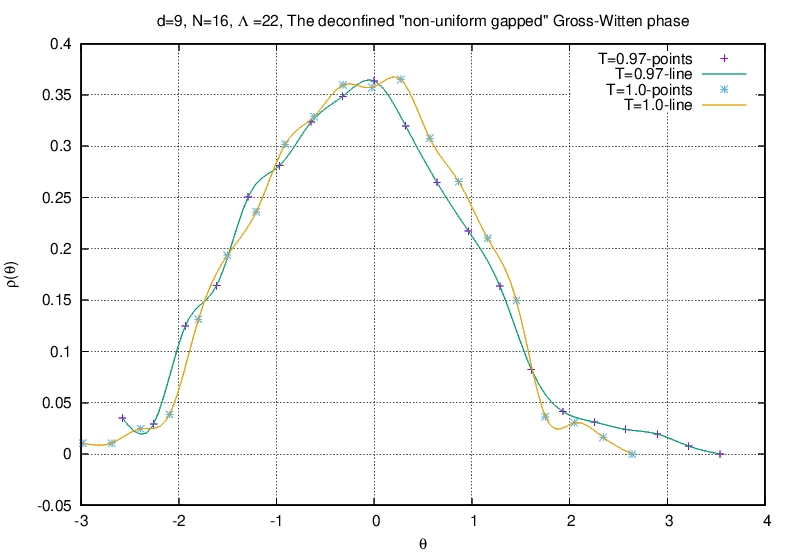}
\captionof{figure}{A plot of the eigenvalues distribution of the holonomy angles $\theta_{a}$ for different temperatures with $N=16$, $d=9$ and $\Lambda=22$.}\label{theta}
\end{center}

Practically, we have used the GWW distributions below to fit our results and obtain with great accuracy the transitions temperatures. We have verified at each plot that this assumption is pretty correct (some details can be found in appendix)

\begin{equation} 
\rho(\theta) = 
\begin{cases} 
\frac{1}{2\pi} & T < T_{c2}.\\ 
\frac{1}{2\pi}(1+\frac{2}{\kappa}\cos \theta) & -\pi \leq \theta \le +\pi~,~\kappa \geq 2~,~T_{c2} \leq T<T_{c1}.\\ 
\frac{2}{\pi\kappa}\cos\frac{\theta}{2} \sqrt{\frac{\kappa}{2}-\sin^2\frac{\theta}{2}} & \theta_0\leq\theta\leq \theta_0~,~\kappa < 2~,~T_{c1} \leq T. 
\end{cases} 
\end{equation}

The plots shown in figure \ref{theta} act more like a visual definition of each phase, we can summarize them as

\begin{enumerate}

\item In the first plot (and in lower temperature region), we may recorgnize that the holonomy angles $\theta_{a}$'s are uniformly distributed in the interval $]-\pi,\pi]$. They correspond to the first GWW distribution $\rho(\theta)=\frac{1}{2\pi}$ and is knowed as the Eguchi-Kawai Phase. The D0-brane are sayed to be \textit{"confined"} here.

\item In the second plot, we observe that the $\theta_{a}$'s act in a different way, it become non-uniformly distributed but is still can take values in the entire interval $]-\pi,\pi]$. This is the non-uniform gapeless GWW phase. Knowed also as a \textit{"partially deconfined phase”} \cite{hanada2019partial}\footnote{Simply, confined and deconfined phases are coexisting in the space of color degrees of freedom, or internal space. Similar coexistence of two phases is very common in usual space, which we denote as "physical space".  For example, at 1-atm and zero-celsius, liquid water and ice coexist \cite{hanada2019partial}.}.

\item In the third plot, we see that the angles are always non-uniform, but a form of a Gap appears, such that some large values are prohibited. It is not shown here, but far from this region (in large temperature), the $\theta_{a}$'s will looks like a Dirac-delta function. This is the non-uniform gapped GWW phase. 
\end{enumerate}

It appears more clearly here that there is two different transitions. The first one is from the uniform $\to$ non-uniform phases: which is the so-called confinement-deconfinement phase transition, or also know as the Hagedorn transition. It is of a second order transition that occurs at a temperature $T_{H}=T_{c2} \approx 0.87$. Sometimes it sayed to be from a confinement to a partial deconfinement phases.

The second one leaves the angles non-uniformly distributed, but generate a gap from a gapeless phase to a gapped phases. It is a third order phase transition know as Gross-Witten-Wadia phase transition that occurs at a temperature $T_{c1} \approx 0.94$. Also called from a partial deconfinement to a complete deconfinement phases.
    
You may look at figure below for a better comparison of theses phases. Our results at this stage is in nice agreement with the work accomplished in \cite{filev2015}.

\begin{center}
\captionsetup{type=figure}
\includegraphics[scale=0.45]{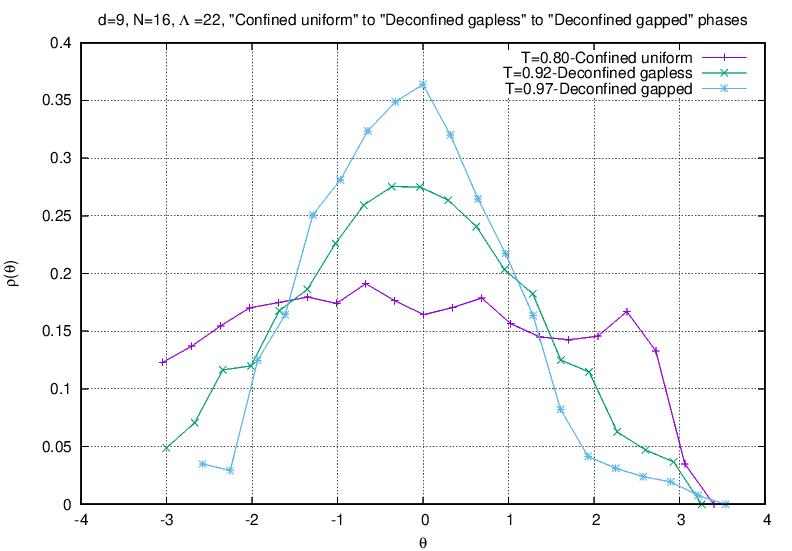}
\captionof{figure}{Different plots of the holonomy angles $\theta_{a}$ to show more explicitely the different phases of the model for $N=16$, $d=9$ and $\Lambda=22$.}
\end{center}

And the three phases are schematically represented in figure \ref{theta-comp}
 
\begin{center}
\captionsetup{type=figure}
\includegraphics[scale=0.55]{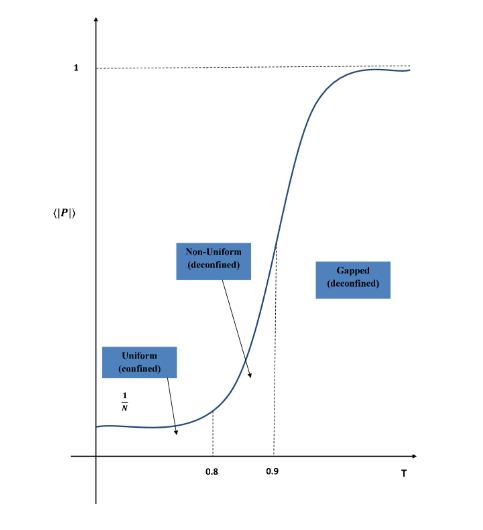}
\captionof{figure}{A schematic representation of the phase structure in terms of the Polyakov line. Taken from \cite{ydri2017review}.}\label{theta-comp}
\end{center} 

\subsection{The hagedorn/Deconfinement transition and the Gravity dual}

One of the most beautiful achievement of physics in last century is the theory of general relitivity (GR) based on Einstein field equations

\begin{eqnarray}
G_{\mu \nu}+\Lambda g_{\mu \nu}=8\pi G T_{\mu \nu}.
\end{eqnarray}

where $G_{\mu\nu}$ is the Einstein tensor, $g_{\mu\nu}$ is the metric tensor, $T_{\mu\nu}$ is the stress–energy tensor and $\Lambda$ is the cosmological constant. Although GR were initially formulated in the context of a 4-dimensional theory where space and time where supposed to be in equal footing, some theorists have explored their consequences in higer dimensions \cite{myers-perry}. Starting from the investigations of Kaluza and Klein that aim to generalize Einstein's ideas to include the electromagnetic force around the idea of a fifth dimension \cite{kaluza,klein}. Laterly these so-called Kaluza-Klein theories did not seem to lead anywhere, and were abandoned. But the subject of higher dimensional theories stay to be of well interest and curiosity\footnote{Don't miss that all what we have done in this thesis are based on 10D superstring theoris in the aim to found an 11D M-theory.}.  

Beside all their success, GR faced two "well-known" space-time singularities where the gravitational field of a celestial body is predicted to become infinite by GR as 1) A singularity inside black holes and 2) A universe singularity, namely the so-called "Big Bang singularity".

Black holes are perhaps the most puzzling objects in general relativity, it is a region of spacetime where gravity is so strong that nothing can escape from their event horizon (the singularity hide behind it). Their study in higher dimension was also largely studied in the context of string theories and are of great interest for us here.

Firstly, a solution of the equations that generalizes a black hole solution in p additional spatial dimensions is knowed are the so-called \textit{black p-brane} and thus a generalization of a black hole is a black p-brane \cite{duff1996black,mohaupt2000black,galtsov2005general}.

Always in higher dimension ($D>4$), a generalization of a black hole solutions that asymptote to a Minkowski space $M^{D-1} \times S^1$ is defined as a \textit{black string}\footnote{Here, the simplest solution one can construct is the uniform black string which is the ($D-1$)-dimensional Schwarzschild-Tangherlini black hole \cite{Tangherlini:1963bw} plus a flat direction, which has horizon topology $S^{D-3} \times S^{1}$.}, the pioneering work on the black string is due to Lemos \cite{lemos1994twodimensional,lemos1994cylindrical} and for a recent and pedagogical overview of black holes in higher dimensions see \cite{emparan2008black}.

In fact, the task is more complex and the study of higher dimensional theory in spacetimes with compact dimensions reveals the existance of several black object solutions\footnote{In this nomenclature, the term “black hole” stands for any object with an event horizon.} including the black-hole and the black-string which are low energy solutions of string theory\footnote{We also allow for the possibility of multiple disconnected event horizons, to which we refer as \textit{multi-black hole solutions}.} \cite{horo-stro}. 

In 1993, R.Gregory and R.Laflamme investigate the evolution of small perturbations around black strings and showed that for a given circle size, uniform strings below a critical mass are linearly unstable \cite{gregory1993black} and it was reexamined through numerical simulation in \cite{unstablestring2003}. Motivated by dynamical considerations \cite{PhysRevLett.87.135001}, Gubser study non-uniform solutions that differ only perturbatively from uniform ones (the one we discuss in last paragraph) and find evidence of a transition from uniform to non-uniform solutions \cite{gubser2001nonuniform}. An interesting property that has been found in this context is the existence of a critical dimension $D^*$ where the transition of the \textit{uniform black string} into a new \textit{non-uniform black string} changes from first order into second order \cite{sorkin2004critical}. Thus, the non-uniform black string phase emerges from the uniform black string phase at the Gregory-Laflamme point.

Moreover, it has been shown \cite{sorkin2006nonuniform,kol2002topology,wiseman2002black,kol2003evidence,kleihaus2007interior} that the localized black hole phase meets the non-uniform black string phase in a horizon-topology changing merger point.  

The problem of these transitions between the 3 branches of solutions - the localised black holes, uniform and non-uniform strings - raises puzzles and addresses fundamental questions such as topology change and it is also an interesting through the Gauge/Gravity correspondence, the Gregory-Laflamme transition in gravity is expected to be related to the phase structure of large N SYM theory compactified on a circle \cite{susskind1998,grav2,aharony2004black,harmark2004new}, and this is why the topic is intrinsically linked to our Monte-Carlo simulation and to matrix models in general in the way we will describe below. To read more about all the technical development we have missed for a desire of briefness, you can refer to \cite{obers2008black,kudoh2004connecting} and references cited above. 

What follow will be based on \cite{kawahara2007phase,string3,ydri2020}. Remind first that the BFSS matrix model is a theory of N coincident D0-branes in type IIA string theory. The fundamental idea from the Gauge/Gravity duality applied to this theory can be written as,

\vspace{1cm}

\begin{center}
\begin{tabular}{lll}
\centering
\textbf{\hspace{1.5cm}Gravity side} && \textbf{\hspace{1.9cm}Gauge side}\\
\textit{The positions of the D0-particles.} &$\longleftrightarrow$& \textit{The eigenvalues of the holonomy matrix.}
\end{tabular} 
\end{center}

\vspace{1cm}

So, the phase of the $\theta_i$'s  is precisely the position of the i-th D0-brane on the circle. If all the angles are distributed uniformly on the circle then we obtain a \textit{uniform black string}, whereas if they accumulate at the same point then we obtain a \textit{black hole} at that location. The phase between them are knowed as a \textit{non-uniform black string phase}. The black hole/black string transition is a very important example of topology change transitions.

As sayed in last section, our model exhibit three phases, see figure \ref{theta} and related talk, viz. 

\begin{enumerate}

\item Thes first phase at low T where the $\theta_i$'s eigenvalues are uniformly distributed in the space, and this correspond in the dual gravity to the uniform black string phase. 

\item The intermediate phase where we have a non-uniform $\theta_i$'s distribution may correspond to be a "little black hole" and not all the D0-brane have been yet concerned with the transition. It is equivalent to to the non-uniform black string.

\item The last one at high T where the $\theta_i$'s eigenvalues distribution are in a gapped phase and became more a more localized are considered to describe a black hole.

\end{enumerate}

Remind that the first transition have been defined as a Confinement/Deconfinement transition, which is a second order phase transition associated with the spontaneous breakdown of the $U(1)$ symmetry

\begin{eqnarray} 
A(t)\longrightarrow A(t)+C.{\bf 1}.  
\end{eqnarray}

At low temperatures this symmetry is unbroken and as a consequence we have a confining  phase characterized by a uniform eigenvalue distribution, i.e.

\begin{eqnarray} 
\rho(\theta)=\frac{1}{2\pi}~,~T\longrightarrow 0.  
\end{eqnarray}

As the temperature increases the above $U(1)$ symmetry gets spontaneously broken and we enter the deconfining phase which is characterized by a non-uniform eigenvalue distribution.

In another side, it has been argued that the deconfinement phase transition in gauge theory such as the above discussed phase transition is precisely the Hagedorn phase\footnote{The Hagedorn temperature was introduced first in the context of hadronic physics. Nowadays, the emphasis is shifted to fundamental strings which might be a necessary ingredient to obtain a consistent theory of black holes \cite{hagedorn1}.} in string theory \cite{aharony2003hagedorn,aharony2004black,hanada2019partial}. Indeed, This corresponds in the dual gravity theory side to a transition between the black hole phase (gapped phase) and the black string phase (the uniform phase) \cite{susskind1998}, the two linked by an intermediate non-uniform black string phase. Thus, the BFSS model at temperature T and large N limit corresponds in the gravity picture to the near-horizon D0-brane solution at finite temperature.

\subsection{Energy and extent of space}

The polyakov line and the Holonomy matrices is not the only results we can get from our model. We present in figure \ref{energy-extent} the plots of the Energy and the Extent of space\footnote{It represents the extent of the eigenvalue distribution of the $\phi_{i}$'s.} against the temperature in the region we are interested in. 

\begin{center}
\captionsetup{type=figure}
\includegraphics[scale=0.5]{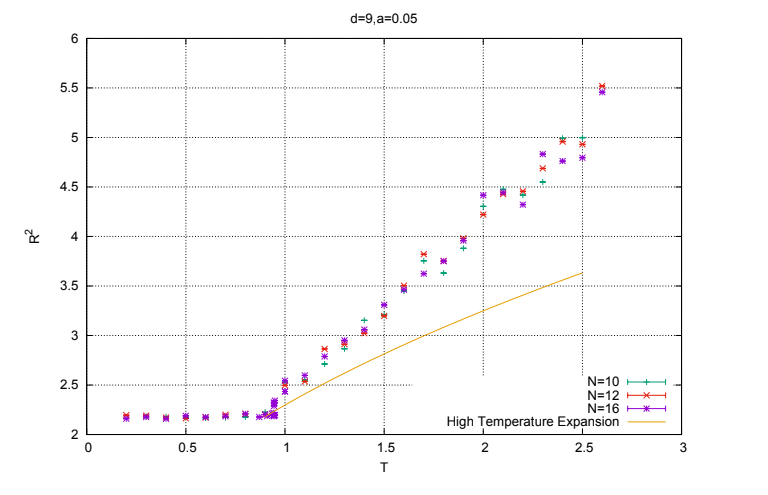}
\includegraphics[scale=0.5]{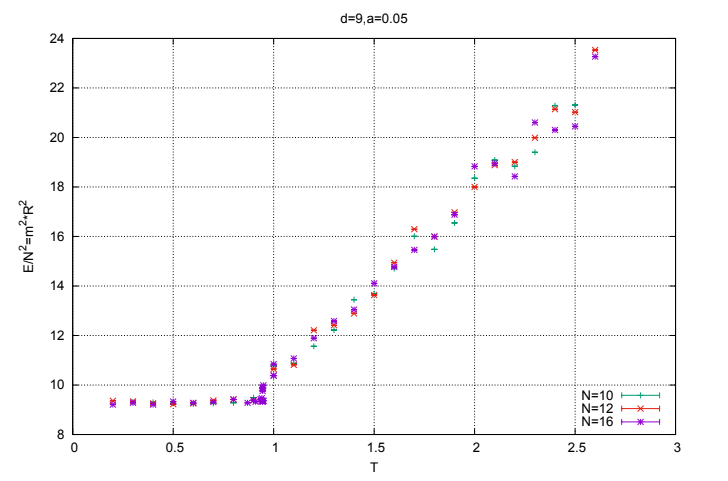}
\captionof{figure}{The figure above show the plots of Energy $E/N^{2}$ and Extent of space $R^{2}$ as functions of the temperature respectively. $N=12, 10, 16$, $d=9$ and $a=0.05$ as before. One can see that near $T \approx 0.9$ the plots suggest the existence of a second order phase transition.}\label{energy-extent}
\end{center}
\vspace{1cm}

As a first remark, the energy and the extent of space show a flat behavior in the confined (uniform) phase for $T < 0.9$. In fact this confirm the relation \eqref{energ} which link the energy and the extent of space by a factor of $m^{2}$. However, even at this point we have excellent overall agreement with the studies of the model in \cite{filev2015}. 

For $T \leq 0.9$ We observe that the results are independent of T, this can be understood as a consequence of the Eguchi-Kawai equivalence \cite{eguchi-kawai}.

The constant value of the energy in the confining uniform phase is identified with the ground state energy. The energy in the deconfining non-uniform phase $(T > T_{c2})$ deviates from this constant value quadratically, i.e.  as $(T-T_{c2})^{2}$. This is confirmed in the full bosonic model \eqref{bosonicBFSS} studied in \cite{kawahara2007phase}.

In this approximation, it is observed that the eigenvalues of the adjoint scalar fields $\Phi_i$ are distributed according to the so-called Wigner semi-circle law.

with a radius $r$ following the temperature behavior of the extent of space $R^{2}$ since $r^2=4R^2/d$, we can resume it into two points, viz.

\begin{enumerate}

\item At low temperature, this radius becomes constant given by $r=\sqrt{2/m}$.

\item Only the radius of the eigenvalue distribution undergoes a phase transition not in its shape (which is always a Wigner semi-circle). 

\end{enumerate}

Recall that the full BFSS matrix model is obtained after a dimensional reduction of $9+1$ dimentional SYM to a $0+1$ theory. In the heart of this model is that the 9 reduced (space-) dimension are now encoded in the 9 adjoint scalar matrices. It is encoded in such a way that the eigenvalues of $\phi^2_{ij}$ correspond in the gravity dual to the radial coordinate of the D0-branes. viz, 

\begin{center}
\captionsetup{type=figure}
\includegraphics[scale=0.4]{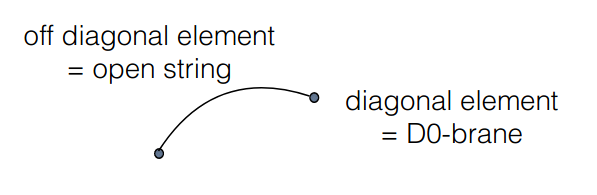}
\captionof{figure}{Classical vision of how we can think about matrix theory work physically. Taken from M.Hanada talk entitled \textit{"What lattice can do  for quantum gravity"}.}\label{diag}
\end{center}
\vspace{1cm}
An let your imagination "play" about what we can construct with this! For example, if the off-diagonal element are suppressed, we get a gas of D0-brane. We can try to describe the Collision of 2 black holes, or the evaporation of it. 

Going back to out last developpement in this chapter, we have to remark that all these plot's have to be seen to describe the same thing from different point of view. Viz, 

\begin{itemize}

\item In the phase below $T_{c2}$, the holonomy angles are \textit{uniformly distributed} around the circle, they are \textit{confined} and distributed according to a Wigner semi circle of constant radius and their energy are also constant and are in the ground state. It is a description of a \textit{uniform black-string}.

\item As we increase the temperature but stay below $T_{c1}$, the holomy angles begin to be excited and regroupe in a more \textit{non-uniformly} localized region (arbitraritly around 0) but stay to cover all the circle $]-\pi,\pi]$, they are \textit{gappeless}. Their energy and radius stay constant but begin to be in a  \textit{deconfinced} phase. This can be interpreted as a "little black-hole" or a \textit{non-uniform black-string}. 

\item As we go to more higher temperatures, the D0-branes localise in a more specific region, they develop a gap $\theta_0$ and are sayed to be \textit{non-uniformly gapped}. The radius conserve their Wigner semi circle property, but the radius it-self increase (they can take more large values) and the energy growth proportionally. It correspond to a \textit{black-hole phase}. 
 
\end{itemize} 

\section{Conclusions}

\begin{itemize}

\item The Gaussian bosonic BFSS matrix model exhibit two transitions:  one from confinement to partial deconfinement, and the other from partial deconfinement to complete deconfinement. These transitions are the Hagedorn transition and Gross-Witten-Wadia (GWW) transition, respectively. They reproduce well the Black-hole/black-string topology change.

\item As argued in \cite{filev2015}, we are confident to say that the dynamics of the bosonic BFSS model is fully dominated by the large d behavior encoded in the quadratic action \eqref{gaussian}. This has been checked in Monte Carlo simulations where a Hagedorn/deconfinement transition is observed consisting of a second order confinement/deconfinement phase transition closely followed by a GWW third order transition.

\item bosons only is not suficient to reproduce the energy of the gravity side (it will be to easy if!) We need to rafinate our result with large N and continuum limit. We can also aim to upgrade our algorithm to contain fermions.

\item Still matrix models have shown a rich benefits for us, we have to find a more complete understanding of them.

\item Ever for the full model, the parameter region relevant for M-theory has not been studied yet, though the parameter region related to IIA string has been studied (focusing on the quantities protected by supersymmetry).
\end{itemize}

\chapter{\emph{Simulation II: Matrix/Geometry approach}}

\epigraph{\textit{Physics is really nothing more than a search for ultimate simplicity, but so far all we have is a kind of elegant messiness.}}{Bill Bryson.}

We have talked (or at least try to...) about the BFSS matrix model, but theorist haven't stop here ! As we have sayed in the introductory section of this thesis, the main goal and holy grail is to found a theory of QUAMTUM GRAVITY !

We will see in this chapter the topic of \textit{emergent geometry}, his origin and some words about Non-commutative geometry and fuzzy physics. Then review a seminal paper that treat how it is possible to get a fuzzy sphere from a BFSS-like model and then we will include a Gaussian approximation (and a regularization) to it. We then move to the phases of the model as a "baby fuzzy sphere" phase and how they are linked to the "Yang-Mills phase", we will also present the behavior of the Hagedorn transition of this model.

As we have tried to be brief as possible, we recommand to consult \cite{ydri2017review} for a mathematical completeness and \cite{ydri2020} to a more complete discussion about our simulation results.

\section{Quantum geometry} 

There is not only the string/M-theory approach to it, but there is also the \textit{Non-commutative geometry} also called \textit{Quantum geometry} approach that we will attemp to explain briefly in this introduction and take it as a continuation of the first one given at the beginning of this thesis.

\subsection*{Non-commutative geometry}

Firtly and to avoid misunderstanding, because the term \textit{Non-commutative geometry} (NCG) is quite ambiguous and people attach to it different meanings. In mathematics, it is an attemp to have a geometrical view on non-commutative algebra in which the multiplication is not commutative, and with the construction of spaces that are locally represented by non-commutative algebras. let's try to give in some points a map to this definition, 

\begin{itemize}
\item The differential geometry of a manifold can be described in terms of an algebra of functions defined on it. The coordinates are the generators of the algebra and vector fields are the derivations.

\item Then, it is natural to attempt to develop a non-commutative version of differential geometry by replacing the algebra of functions $\mathbb C$ by an abstract associative algebra $\mathbb A$ which is not necessarily commutative.
\end{itemize}

The idea of noncommutative geometry is to encode everything about the geometry of a space algebraically and then allow all commutative function algebras to be generalized to possibly non-commutative algebras.

One of the main difference is the loss of the idea of localization. There is no longer a well defined notion of a point. In contrary of what we can think, this is a strong point of the theory !

And what a beautiful concept of that of a point in Natural Science ! How easy is to say where objects are when one has introduced such a precise definition of localization ! And then laws were found for the interactions of moving bodies and geometries to describe them (the best known example is the one of gravitation).

Even with classical mechanics (in the canonical formulation), we deals with non-commutative structure. Namely, the Poisson brackets where q and p are the habitual generalized coordinates and momenta, 

\begin{eqnarray}
\{f,g\}_{\{q,p\}}~=~\frac{\partial f}{\partial q}\frac{\partial g}{\partial p}~-~\frac{\partial f}{\partial p}\frac{\partial g}{\partial q}~,~\{p,q\}_{\{q,p\}}~=~-1.
\end{eqnarray} 

Now when we move to quantum mechanics, where we are no more allowed to say neither where an electron is exactly located on its "orbit" around the proton in the hydrogen atom nor to describe the trajectory of the photon in the Young’s double slit experimen. Where the main feature of quantum mechanics is the non-commutativity of observables as,

\begin{eqnarray}
\{\hat{p},\hat{q}\}~=~-i\hslash
\end{eqnarray}

Analogously, in NCG the coordinates fulfill the "canonical" relation as,

\begin{eqnarray}
[\hat{x}^\mu,\hat{x}^\nu]~=~i\Theta^{\mu\nu}.\label{13}
\end{eqnarray}  

which means that (with any given set of axes), it is impossible to accurately measure the position of a particle with respect to more than one axis. In fact, this leads to an uncertainty relation for the coordinates analogous to the Heisenberg uncertainty principle. The equation \ref{13} leads to the so-called \textit{Groenewold-Moyal-Weyl} product or the $\star$-product. 

\subsection*{Fuzzy sphere}

A very concrete example of a non-commutative space is the \textit{Fuzzy sphere} space \cite{madore1992fuzzy}. A fuzzy sphere is a variant of an n-sphere in non commutative geometry. Often the fuzzy 2-sphere is meant by default, but there are also fuzzy spheres of higher dimension.

Fuzzy spaces lead to matrix models too and their ability to reflect topology better than elsewhere should therefore evoke our curiosity \cite{balachandran2005lectures}.

Fuzzy physics is in the heart of this chapter, we will skip some details given in \cite{balachandran2005lectures,balachandran2006noncommutative,ydri2017review}. Take in mind that fuzzy spheres appear as classical solutions in the pp-wave matrix model \cite{berenstein2003strings} which is a generalization of the BFSS matrix theory to the so-called pp-wave background \cite{KOWALSKIGLIKMAN1984194} (A pp-wave spacetime is a spacetime containing nothing but radiation).

Last, but not least, nature seems often more subtle than human mind. We mention that there have been some speculations that string theory might give rise naturally to space-time uncertainty relations and then relate it to non-commutative geometry \cite{li1999short}, It might also give rise to a non-commutative theory of gravity (also from Ads/Cft correspondance) \cite{jevicki1999non}. More especially, there have been attempts to relate a non-commutative structure of space-time to the quantization of open strings ...\cite{dubois1989gauge,schomerus1999d}.

\section{The Chern-Simon term}

We will review firstly previous seminal work to relate the bosonic BFSS model to fuzzy physics, then include our touch which consist of a Gaussian approximation as we did in last chapter and we see what will happen.
 
\subsection{Starting point}

Firstly, we will take as a starting point the leading work \cite{kawahara2007fuzzy} where they begun with the bosonic BFSS model \eqref{bosonicBFSS} but here for $d=3$ (three matrices $\phi_{a}$ at each lattice point $\equiv$ namely a BFSS bosonic-like model),

\begin{eqnarray} 
S =\frac{1}{g^2}\int_0^{\beta}dt{\rm Tr}\bigg[ \frac{1}{2}(D_t\Phi_i)^2-\frac{1}{4} [\Phi_i,\Phi_j]^2\bigg].\label{original}
\end{eqnarray}

And then add to it a Chern-Simons term as,

\begin{eqnarray} 
S =\frac{1}{g^2}\int_0^{\beta}dt{\rm Tr}\bigg[ \frac{1}{2}(D_t\Phi_i)^2-\frac{1}{4} [\Phi_i,\Phi_j]^2+\frac{2i\alpha}{3}\sum_{a,b,c=1}^3 \epsilon_{abc} \Phi_a \Phi_b  \Phi_c \bigg].\label{chern}
\end{eqnarray}

Where the cubic term represents the so-called \textit{Chern-Simons term}, which is well known to be responsible for the emergence of the classical geometry of a round sphere ("Fuzzy" because of the non-trivial commutation relation among $\Phi_i$).

Always in \cite{kawahara2007fuzzy}, various thermodynamical quantities\footnote{In addition to previous studied quantities such as the Polyakov line and the radius, there is also a Chern-Simon measure (CS) that will be defined later.} of the model \eqref{chern} have been studied in the large-N limit by a Monte Carlo simulation. We will just look at their phase diagram\footnote{It is important to note that the parameters of the model is not directly $\alpha$ and $T$, but are actually scalled such that $\tilde{\alpha}~=~N^{1/3}\alpha$ and $\tilde{T}~=~N^{-2/3}T$. This comes from perturbative calculations around the fuzzy sphere detailed in \cite{kawahara2007fuzzy}.},

\begin{center}
\captionsetup{type=figure}
\includegraphics[scale=0.63]{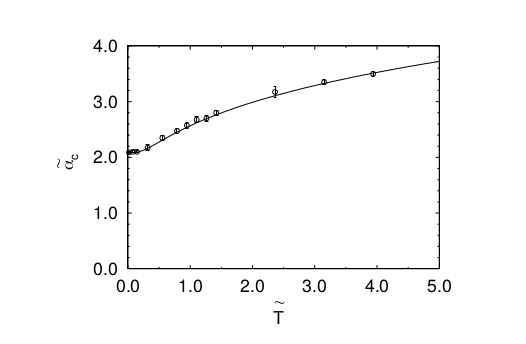}
\captionof{figure}{The critical $\tilde{\alpha}$ plotted against $\tilde{T}$ by a Monte-Carlo simulation for N = 16. Taken from \cite{kawahara2007fuzzy}.}\label{phase-diag}
\end{center}

Above the critical line (high coupling constant $\tilde{\alpha}$), it is a phase defined as a fuzzy sphere phase which decays (evaporate) below a critical $\tilde{\alpha}$ that depend on the the temperature $\tilde{T}$ as shown in figure \ref{phase-diag} to a Yang-Mills phase where the vacuum state is given by commuting matrices (also called a matrix phases).

\subsection{The model: A Gaussian approximation + regularization}

As we did in last chapter, wa wanting first to perform a Gaussian approximation (A large d approximation, assuming $d~=~3$ is sufficiently large) gives the following action, 

\begin{eqnarray} 
S =\frac{1}{g^2}\int_0^{\beta}dt{\rm Tr}\bigg[ \frac{1}{2}(D_t\Phi_i)^2+\frac{1}{2}m^2\Phi_i^2+\frac{2i\alpha}{3}\sum_{a,b,c=1}^3 \epsilon_{abc} \Phi_a \Phi_b  \Phi_c \bigg]~,~m~=~d^{1/3}.\label{chern1}
\end{eqnarray}

Unfortunately, it was not too simple. Even if the action \eqref{chern1} thermalize normally, their non-perturbative physics that should be accessible by a Monte-Carlo simulation still "hidden" because it not bounded from below and that the solution (fuzzy solution) are not so stable in this action. 

We were not discouraged from that and inspired by a work done in \cite{delgadilo} (section 4.8), we regularize  it (embed it)  by  adding  a  quartic  potential term to the previous action as, 

\begin{eqnarray}
S_{R} = M \int_0^{\beta}dt\bigg({\rm Tr}\Phi_i - \alpha c_{2}N \bigg)~,~c_{2}~=~\frac{N^{2}-1}{4}.
\end{eqnarray}

Readers interested in this step are encouraged to consult section 3.1 of \cite{ydri2020}.

Assuming the scaling of $\alpha$ and $T$ to be identical to the scaling in the exact model (with the full Yang-Mills term instead of the harmonic oscillator term with $m~=~d^{1/3}$),

\begin{eqnarray}
\tilde{\alpha}=N^{1/3}\alpha ~,~\tilde{T}=N^{-2/3}T
\end{eqnarray}

We can then move to study the different phases of the reularized model.

\section{Yang-Mills phase and Wigner semi-circle law}

Starting from the following two assumptions, 

\begin{enumerate}
\item The  Chern-Simons  term  vanishes  in  the  Yang-Mills  phase, indicating the dominance of commuting and diagonal matrices.

\item The three commuting matrices are static and by rotational invariance their contributions are identical.
\end{enumerate}

The eigenvalue distribution $\rho(\lambda)$ in the Yang-Mills phase are effectively given by a Wigner semi-circle law, 

\begin{eqnarray}
\rho(\lambda)~=~\frac{2}{\pi R^2}\sqrt{R^2 - \lambda^{2}}.\label{rho}
\end{eqnarray}

Where there are two quite different behaviours depending on the values of the coupling $\tilde{\alpha}$ such that\footnote{Here we summarize the results, details can be found in \cite{ydri2020}.}, (it is important to note that the mass deformation M is taken to be large)

\begin{enumerate}

\item For $\tilde{\alpha}~\neq ~0$ corresponding to commuting matrices where the radius in the Wigen semi-circle law is given by (for $d~=~3$),

\begin{eqnarray}
R ~=~ \frac{\tilde{\alpha}N^{2/3}}{\sqrt{d}}+\mathcal{O}(\frac{1}{M}).\label{behave1}
\end{eqnarray}

Which are independent of M and $\tilde{T}$, but depend linearly on $\tilde{\alpha}$ and scales as $N^{2/3}$ as shown in the Monte-Carlo measure in figure \ref{yang1} for $\tilde{\alpha}~=~0.2$, 

\begin{center}
\captionsetup{type=figure}
\includegraphics[scale=0.61]{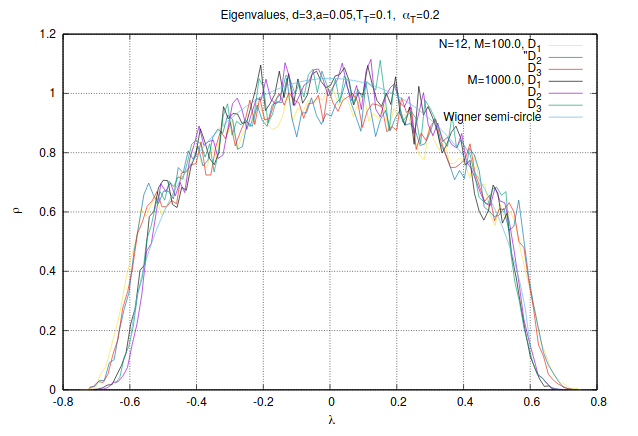}
\captionof{figure}{The first behavior of the Wigner semi-circle law as in \eqref{behave1} for $N~=~12$.}\label{yang1}
\end{center}

\item For $\tilde{\alpha}~=~0$ corresponding to the minimum $\Phi_a~=~0$ exactly, the Wigner semi-circle radius is then given by (for $d~=~3$), 
\begin{eqnarray}
R &=& \bigg(\frac{4 T}{d^2 M}\bigg)^{1/4}.\label{behave2}
\end{eqnarray}

Where we see clearly the T-dependence of the Wigner semi-circle radius in this casen the Monte-Carlo measure of this at $\tilde{\alpha}~=~0,1$ is given in fugure \ref{yang2},

\begin{center}
\captionsetup{type=figure}
\includegraphics[scale=0.6]{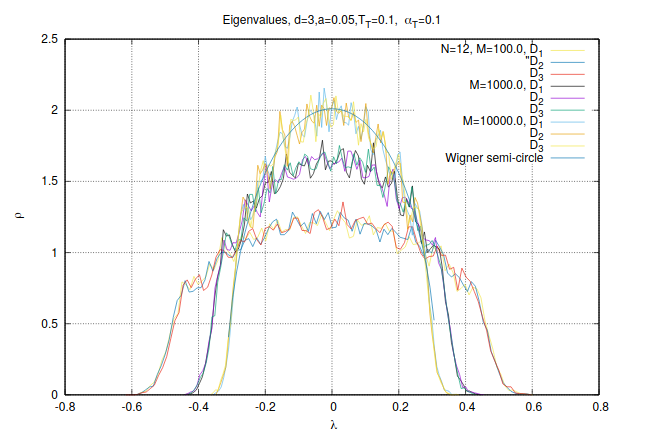}
\captionof{figure}{The second behavior of the Wigner semi-circle law as in \eqref{behave2} for $N~=~12$.}\label{yang2}
\end{center}

\end{enumerate}

\section{Other observables}

There is not only the $\Phi_a$ eigenvalues. As before, there is also the Polyakov line $\langle |P|\rangle$, the Radius $R^2$ and in addition there is the Chern-Simon (CS) observable respectively,

\begin{eqnarray} 
\langle |P|\rangle =\frac{1}{N}\sum_a\exp(i\theta_a).\label{1}
\end{eqnarray}

\begin{eqnarray} 
R^2=\frac{1}{N\Lambda}\sum_{n=1}^{\Lambda}{Tr}{\Phi}_i^{2}(n).\label{2}
\end{eqnarray}

\begin{eqnarray}
CS=\frac{1}{N\Lambda}<\frac{2i}{3}\epsilon_{abc}\sum_{n=1}^\Lambda\Phi_a(n)\Phi_b(n)\Phi_c(n)>.\label{3}
\end{eqnarray}

\begin{figure}[!ht]
\captionsetup{type=figure}
\centering
    \begin{subfigure}[b]{0.47\textwidth}            
            \includegraphics[width=\textwidth]{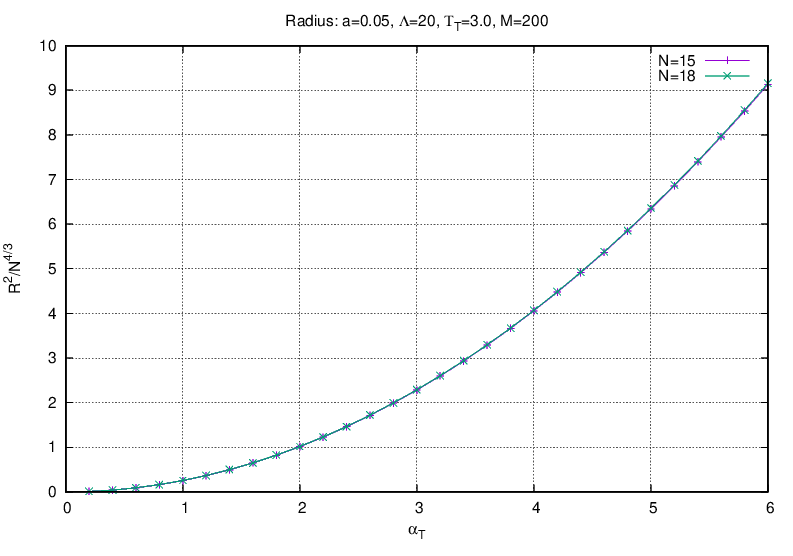}
    \end{subfigure}%
    ~
    \begin{subfigure}[b]{0.47\textwidth}
            \centering
            \includegraphics[width=\textwidth]{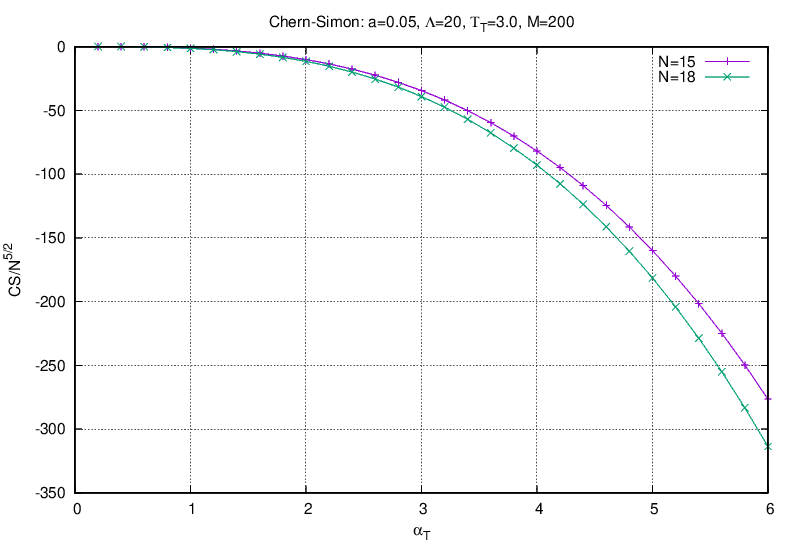}
    \end{subfigure}
    \caption{The observables $R^2$ (left) and CS (right) for $\tilde{T}=3.0$.}\label{observale-t}
\end{figure}

\begin{figure}[!ht]
\captionsetup{type=figure}
\centering
    \begin{subfigure}[b]{0.49\textwidth}            
            \includegraphics[width=\textwidth]{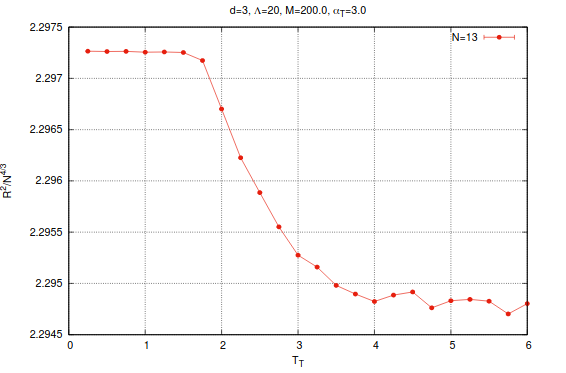}
    \end{subfigure}%
    ~
    \begin{subfigure}[b]{0.495\textwidth}
            \centering
            \includegraphics[width=\textwidth]{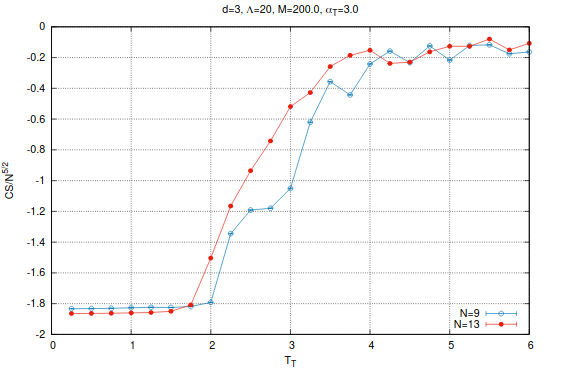}
    \end{subfigure}
    \caption{The observables $R^2$ (left) and CS (right) for $\tilde{\alpha}=3.0$.}\label{observale-a}
\end{figure}

A closer look is given to the Polyakov line in a distinct section at the end of the chapter (In the Hagedorn transition discussion). For now let's deal with the radius \eqref{2} and Chern-Simon \eqref{3} observables. For instance, we have put in figure \ref{observale-t} a sample of them for $\tilde{T}=3.0$ by varying $\tilde{\alpha}$ and in \ref{observale-a} for fixed $\tilde{\alpha}=3.0$ and varying $\tilde{T}$.

Where we can clearly decompose the graphs into two region, a first region with constant value for the radius (left) as well as for the Chern-Simon (right) where we clearly move to the second region. The first region correspond to the Yang-Mills phase we talked about before and the second region is another phase we will talk about in the next section in term of the eigenvalues distribution as we did in the first phase. This is the same shape as for the full model \eqref{chern}.

\section{The "Baby" fuzzy sphere phase}

We've seen from the observables \eqref{1} and \eqref{2} that at certain values of the coupling contant $\tilde{\alpha}$ for contant $\tilde{T}$ \ref{observale-t} (or inversly \ref{observale-a}) their comportement change, this suggest a transition from one state to another: a phase transition. 

We have presented the first phase (the Yang-Mills phase) in term of the eigenvalues distribution which are given theoretically by the Wigner semi-circle law \eqref{rho}, even if we have the two different behaviour \eqref{behave1} and \eqref{behave2} for high temperatures. Let's now discover the behaviour of the eigenvalues distribution and the second phase when we increase the value of the coupling constant in figure \ref{fuzzy},

\begin{center}
\captionsetup{type=figure}
\includegraphics[scale=0.69]{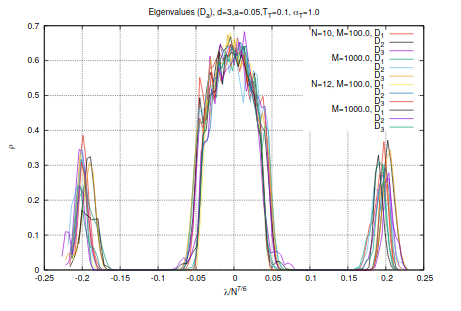}
\captionof{figure}{The baby fuzzy sphere configurations for $\tilde{T}=0.1$ and $\tilde{\alpha}=1.0$ for various N and M.}\label{fuzzy}
\end{center}

This is a fuzzy-sphere-like phase because it corresponds in some sense to a fuzzy sphere with three cuts only. This three-cut phase is termed in this thesis as a "baby fuzzy sphere phase" because unlike the Yang-Mills phase it is a geometric phase with all the characteristics of the fuzzy sphere phase, here we have judged important to give you a view on what a real fuzzy sphere looks like in term of eigenvalues distribution in figure \ref{fuz}\footnote{The figure are taken from \cite{delgadilo} and correspond to the eigenvalue distribution of one of the three matrices for the matrix model (3.6) in the cited article and it write as, 
\begin{eqnarray}
S~=~\frac{\tilde{\alpha}^4}{N}{\rm Tr}\bigg[-\frac{1}{4}[D_a,D_b]^2+\frac{2i}{3}\epsilon_{abc} D_a D_b D_c +\frac{m^2}{2c_2}(D_a^2)^2-\mu D_a^2\bigg].\nonumber
\end{eqnarray} 
},

\begin{center}
\captionsetup{type=figure}
\includegraphics[scale=0.64]{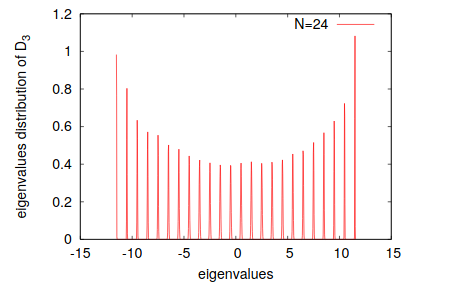}
\captionof{figure}{Eigenvalue distribution in the fuzzy sphere phase. Taken from \cite{delgadilo}-Figure 16.}\label{fuz}
\end{center}

We say that there is an emergence of a geometry. This is the geometry of the sphere emerging from a matrix model (as the one observed in the original exact model \eqref{original}). For our actual model $S+S_R$, we only capture a remnant of this phase. As we lower the coupling, a transition will occur to a one-cut phase (the Yang-Mills phase) where all eigenvalues are centered around $\lambda=0$ and are continuous. We will try to schematically resume all that by the phase diagram of the model in next section.

\section{The phase diagram}

We have not derived the phase diagram directly from the matrix model we studied, we are simply spelling out what answer we expect in terms of the two parameters of the model, $\tilde{\alpha}$ as a function of $\tilde{T}$,
\vspace{1cm}
\begin{center}
\captionsetup{type=figure}
\includegraphics[scale=0.53]{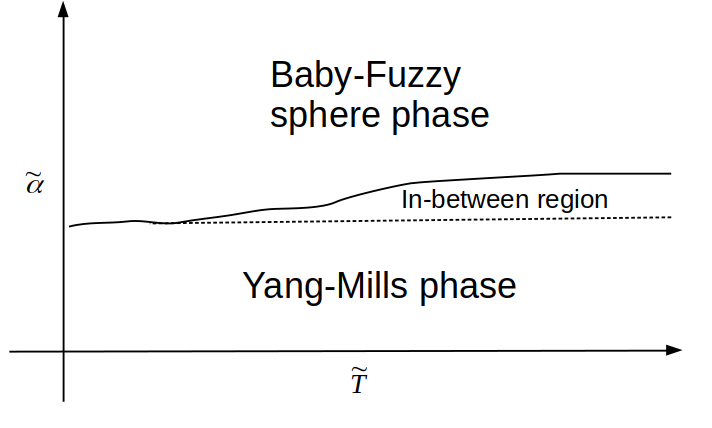}
\captionof{figure}{The expected phase diagram of the regularized model where we observe the Yang-Mills phase at small coupling and the baby-fuzzy sphere phase at high coupling, the in-between region (which are a part of the yang-Mills phase) is showed for high temperatures.}
\end{center}
\vspace{0.5cm}
We have identified two different phases of the matrix model, 

\begin{enumerate}
\item The Yang-Mills phase that are divided into two distinct regions. The first region is dominated by the Wigner’s semi-circle law at very low values of the gauge coupling constant and a second region dominated by a uniform distribution occurring at medium values of the gauge coupling constant before reaching the baby fuzzy sphere boundary.

\item The boundary between the two phases is constructed and it is shown that the scaling of the gauge coupling constant and the temperature in the two phases is possibly different resulting in the fact that either the geometric baby fuzzy sphere phase removes the uniform confining phase (which is the most plausible physical possibility). In the matrix phase the fuzzy sphere vacuum collapses under quantum fluctuations.
\end{enumerate}
\cleardoublepage
\section{Hagedorn transition again}

In the previous chapter, we see that the studied system (where the Chern-Simons term is set to zero and $d~=~9$) was characterized by a Hagedorn transition as a confinement phase at low temperatures to a deconfinement phase at high temperatures. This section aim to show that the regularized action $S+S_R$ also exhibit a hagedorn transition and it seems to depend on the radius of the Wigner’s semi-circle law, see the Monte-Carlo results plots shown in figure \ref{hag} to be convinced by that

\vspace{1cm}

\begin{center}
\captionsetup{type=figure}
\includegraphics[scale=0.6]{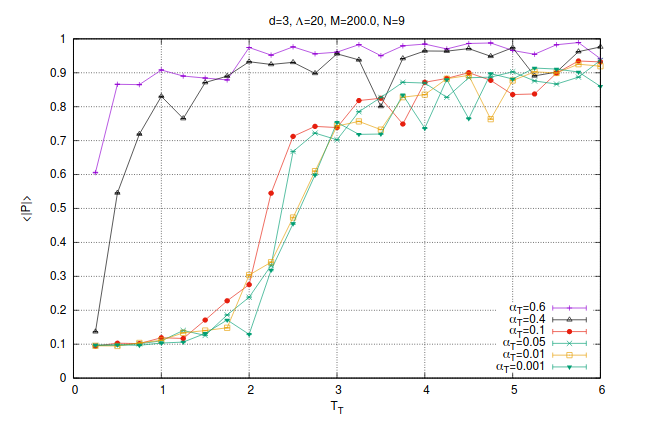}
\captionof{figure}{The Hagedorn transition and the behavior dependence of the polyakov line according to different values of the gauge coupling $\tilde{\alpha}$.}\label{hag}
\end{center}

The figure can be understood into two point, 

\begin{enumerate}

\item At very low values of the gauge coupling constant $\tilde{\alpha}~<~0.1$ where the radius of the Wigner’s semi-circle law is given by \eqref{behave2}, 

\begin{eqnarray}
R = \bigg(\frac{4 T}{d^2 M}\bigg)^{1/4}.\nonumber
\end{eqnarray}

the Chern-Simon term doesn't influate a lot on the Hagedorn temperature and stay to occurs at the same value (arround $\approx 1.7-1.8$) for all values of $\tilde{\alpha}$.

\item For larger values of the gauge coupling constant $\tilde{\alpha}$ where the radius of the Wigner’s semi-circle law is given by \eqref{behave1},

\begin{eqnarray}
R = \frac{\tilde{\alpha}N^{2/3}}{\sqrt{d}}+\mathcal{O}(\frac{1}{M}).\nonumber
\end{eqnarray}

the Hagedorn temperature decreases with increasing $\tilde{\alpha}$ until the confinement phase disappears. Indeed, the baby fuzzy sphere phase exists always in the deconfinement phase.

\end{enumerate}

\section{Conclusions}

\begin{itemize}

\item In this second model, the Yang-Mills phase becomes divided into two distinct regions; 1) The first region is dominated by the Wigner’s semi-circle law at very low values of the gauge coupling constant $\tilde{\alpha}$, 2) A second region inside the Yang-Mills phase dominated by a uniform distribution occurring at medium values of the gauge coupling constant and seems to be a crossover region before reaching the baby fuzzy sphere phase for large values of $\tilde{\alpha}$.

\item The Hagedorn temperature seems to depend on the gauge coupling constant $\tilde{\alpha}$ and as a result of that is that the confinement phase disappears for $\tilde{\alpha}~>~0.1$ but the baby fuzzy sphere phase stay to exists always in the deconfinement phase.

\item The characteristics of the black-hole-to-black-string phase transition are captured to a verygood accuracy using the Gaussian model whereas only a remnant of the Yang-Mills-to-fuzzy-sphere phase is reproduced by means of the cubic action (regularized effectively by a double-trace potential which removes only the zero mode of the scalar sector).

\item The Gaussian approximation of this second model (If we forget out regularization treatement) is certainly valid for small values of $\alpha$ but the extrapolation to much larger values does not necessarily need to be valid ! In fact it is expected to break down at some point.

\end{itemize}

\appendix

\chapter{\emph{Field and angle Variations}}

This appendix regroup the variations needed to run the metropolis algorithm correctly\footnote{at least for the first code, the second one have been constructed in the same manner.} which in the heart of our code. Recall that we are interested in the total lattice action 

\begin{eqnarray} 
S_{\rm total}&=&N\sum_{n=1}^{\Lambda}{
 Tr}\bigg[\frac{1}{a}{\Phi}_i^{ 
2}(n)+\frac{1}{2}am^2{\Phi}_i^{ 2}(n)\bigg]\nonumber\\ 
&-&\frac{N}{a}\sum_{n=1}^{\Lambda}{Tr}{\Phi}_i^{}(n)D_{\Lambda}{\Phi}_i^{}(n+1)D_{\Lambda}^{\dagger}-\frac{1}{2}\sum_{a\ne b}\ln\sin^2\frac{\theta_a-\theta_b}{2}.
\end{eqnarray}

\section{The variation due to bosonic matrices}

We choose the update

\begin{eqnarray} 
(\delta\Phi_i(n))_{pq}=h\delta_{pa}\delta_{qb}+h^*\delta_{qa}\delta_{pb}.  
\end{eqnarray}

The variation is obtained such that we compute $\Delta S_{i}^{\rm total}(n)=S[\Phi_i(n)+\delta\Phi_i(n)]-S[\Phi_i(n)]$,

\begin{align} 
\Delta S_{i}^{\rm total}(n)_{ab}&=\frac{N}{a}\bigg[4(1+\frac{m^2a^2}{2}){\rm Re}(h^*\phi_i(n))_{ab}\nonumber\\
&-2{\rm Re}(f^*\Phi_i(n+1)+g^*\Phi_i(n-1))_{ab}\nonumber\\
&+(1+\frac{m^2a^2}{2})(2hh^*+(h^2+h^{*2})\delta_{ab})\bigg].
\end{align}

where $f^*$ and $g^*$ are given by

\begin{eqnarray} 
f^*=h^*\exp(\frac{i}{\Lambda}(\theta_a-\theta_b))~,~g^*=h^*\exp(-\frac{i}{\Lambda}(\theta_a-\theta_b))  
\end{eqnarray}

\section{Variation due to holonomy angles}

There is also the variation under the change of the angle $\theta_c$ of the holonomy matrix, viz.

\begin{eqnarray} 
\theta_c\longrightarrow\theta_c^{\prime}=\theta_c+\alpha.\label{theta_var} 
\end{eqnarray}

The relevant action here is given by

\begin{eqnarray} 
S(\theta)&=&-\frac{N}{a}\sum_{n=1}^{\Lambda}\sum_{a,b}e^{-\frac{i}{\Lambda}(\theta_a-\theta_b)}({\Phi}_i(n))_{ab}({\Phi}_i(n+1))_{ba}\nonumber\\ 
&-&\frac{1}{2}\sum_{a\ne
 b}\ln\sin^2\frac{\theta_a-\theta_b}{2}. 
\end{eqnarray}

As before, the variation of this action after the update \eqref{theta_var} is given explicitly by

\begin{eqnarray} 
\Delta S_c(\theta)&=&-\frac{2N}{a}{\rm Re}\bigg(\sum_{n=1}^{\Lambda}\sum_{a\ne c}e^{\frac{i}{\Lambda}(\theta_a-\theta_c)}(e^{-i\frac{\alpha}{\Lambda}}-1)({\Phi}_i(n+1))_{ac}({\Phi}_i(n))_{ca}\bigg)\nonumber\\ 
&-&\sum_{a\ne c}\ln\sin^2\frac{\theta_a-\theta_c-\alpha}{2}+\sum_{a\ne c}\ln\sin^2\frac{\theta_a-\theta_c}{2}.
\end{eqnarray}

An extremely important remark is now in order. The action $S(\theta)$ does not depend on the center of mass $\theta_{\rm cm}=\sum_{a=1}^N\theta_a/N$.  Indeed, the functional integration over $\theta_a$ is (almost) identical to the functional integration over $\tilde{\theta}_a=\theta_a-\theta_{\rm cm}$ which satisfies $\sum_{a}\tilde{\theta}_a=0$.

Furthermore, by means of the $U(1)$ gauge transformation $A(t)\longrightarrow A(t) +C .{\bf 1}$ we can choose the static gauge, 

\begin{eqnarray}
A(t)=-{\rm diag}(\theta_1,...,\theta_N)/\beta.
\end{eqnarray}

in such a way that (see \cite{anagnostopoulos2007})

\begin{eqnarray}
max(\tilde{\theta}_a)-min(\tilde{\theta}_a) < 2\pi. 
\end{eqnarray}

We have then explicitly

\begin{eqnarray} 
\int d\theta F(\Delta\theta)=\int d\tilde{\theta}\delta(\sum_a\tilde{\theta}_a)F(\Delta\tilde{\theta})\int d\theta_{\rm cm}. 
\end{eqnarray}

Since $\theta=\tilde{\theta}+\theta_{\rm cm}$ and $-\pi < \theta \leq +\pi$ while ${\rm min}(\tilde{\theta}_a)\leq \tilde{\theta} \leq {\rm max} (\tilde{\theta}_a)$ we conclude immediately that the center of mass $\theta_{\rm cm}$ must be in the interval $]-\pi - {\rm max}(\tilde{\theta}_a),\pi - {\rm min}(\tilde{\theta}_a)]$. Hence the above integral becomes (with $\mu={\rm max}(\tilde{\theta}_a)-{\rm min}(\tilde{\theta}_a))$

\begin{eqnarray} 
\int d\theta F(\Delta\theta)=\int d\tilde{\theta}\delta(\sum_a\tilde{\theta}_a)F(\Delta\tilde{\theta})(2\pi+\mu). 
\end{eqnarray}

Clearly, this is true as long as $\mu < 2\pi$ while for $\mu > 2\pi$ the additional Boltzmann weight is identically zero. We have then the extra  Boltzmann weight \cite{hanada-code}

\begin{eqnarray} 
&&w(\mu)=2\pi+\mu~,~\mu\le 2\pi\nonumber\\ 
&&w(\mu)=0~,~\mu > 2\pi.
\end{eqnarray}

In other words, we can replace the functional integration over $\theta_a$ with the functional integration over $\tilde{\theta}_a$ with an additional Boltzmann weight $w(\mu)$.

Thus, the update $(\ref{theta_var})$ should be thought as $\tilde{\theta}_c\longrightarrow \theta_c^{\prime}=\tilde{\theta}_c+\alpha$ where $\alpha$ is uniformly distributed in the range $]-\pi-{\rm max}(\tilde{\theta}_a),\pi-{\rm min}(\tilde{\theta}_a)]$.

\chapter{\emph{Gross-Witten-Wadia transition}}

We will review in this appendix in a very compact way, few things about the Gross-Witten-Wadia phase transition (GWW). Our goal is not to fully explain it, but to present them in a general picture of it (a detailed study of this subjects is beyond our current goal).

This is a third order phase transition occurring at a temperature $T_{c1} > T_{c2}$ dividing therefore the non-uniform phase into two distinct phases: The gapless phase in the intermediate region $T_{c2} \leq T < T_{c1}$ and the gapped phase at high temperatures $T \geq T_{c1}$.

It is observed  in numerical simulations \cite{kawahara2007phase} that this phase transition is well described by the Gross-Witten-Wadia one-plaquette model given explicitly by \cite{GWW1,GWW2}

\begin{eqnarray} 
Z_{GWW}=\int dU \exp\left(\frac{N}{\kappa}(Tr U+ Tr \dagger{U})\right).  
\end{eqnarray}

This model act as a phenomenological model for the holonomy matrix. The deconfined non-uniform gapless phase is described by a gapless eigenvalue distribution (and hence the name: gapless phase) of the form

\begin{eqnarray} 
\rho_{\rm gapless}=\frac{1}{2\pi}(1+\frac{2}{\kappa}\cos \theta),-\pi \leq \theta \le +\pi ,\kappa \geq 2.
\end{eqnarray}

There is also a deconfined non-uniform gapped phase which is described by the eigenvalues distribution as 

\begin{eqnarray} 
\rho_{\rm gapped}=\frac{2}{\pi\kappa}\cos\frac{\theta}{2} \sqrt{\frac{\kappa}{2}-\sin^2\frac{\theta}{2}}~,~ -\theta_0\leq\theta\leq +\theta_0~,~\kappa < 2.
\end{eqnarray}

Thus, there exists in the GWW one-plaquette model a phase transition between the above two solution  occurring at $\kappa = 2$  which  is  found  to  be  of  third order.

The fact that the angle $\theta$ takes values in the full range $]-\pi,+\pi]$ is precisely what is meant by the word "gapless", i.e. there are no gaps on the circle. This solution is valid only for $\kappa\geq 2$ where $\kappa$ is a function of the temperature. At $\kappa=2$ (corresponding to $T=T_{c1}$) a third order phase transition occurs to a gapped eigenvalue distribution.

At high temperatures corresponding to $\kappa\longrightarrow 0$, the gapped distribution since only the interval $[-\theta_0,\theta_0]$ is filled, the above distribution approaches a delta function \cite{aharony2003hagedorn}.

This third order phase transition is associated therefore with the appearance of a gap in the eigenvalue distribution. We notice that the deconfining non-uniform phase is dominated by the gapped phase since the region of the gapless phase is extremely narrow.

Thus, the Polyakov line suffers another phase transition in the non-uniform phase where it rises from 0 to the value 1/2 at $\kappa=2~(T=T_{c1})$ in the gapless phase then rises further from 1/2 to the value 1 in the gapped phase. We can calculate explicitly their behavior as

\begin{eqnarray} 
P=1-\frac{\kappa}{4}~,~\kappa < 2. \label{eq1}
\end{eqnarray}

\begin{eqnarray} 
P=\frac{1}{\kappa}~,~\kappa \geq 2.   \label{eq2}
\end{eqnarray}

which crosses 1/2 at the critical point $\kappa = 2$. Note that \eqref{eq1} and \eqref{eq2} and its first derivative with respect to $\kappa$ is continuous at $\kappa = 2$, but the second derivative has a discontinuity. Thus, the Gross-Witten model undergoes a third order phase transition at $\kappa = 2$. Practically, we have used these distributions to fit our results and obtain with great accuracy the transition temperature as, 

\begin{eqnarray} 
T_{c1} \approx 0.94
\end{eqnarray}

\cleardoublepage
\phantomsection
\addcontentsline{toc}{chapter}{Bibliography}
\bibliography{biblio} 

\begin{thebibliography}{100}

\bibitem{pais:1988}
A.~Pais, {\em \textit{Inward Bound: Of Matter and Forces in the Physical
  World}}.
\newblock Oxford University Press, USA, 1988.

\bibitem{string0}
D.~Rickles, {\em A Brief History of String Theory: From Dual Models to
  M-Theory}.
\newblock The Frontiers Collection, Springer-Verlag Berlin Heidelberg, 1~ed.,
  2014.

\bibitem{Regge1959IntroductionTC}
T.~Regge, ``Introduction to complex orbital momenta,'' {\em Il Nuovo Cimento
  (1955-1965)}, vol.~14, pp.~951--976, 1959.

\bibitem{PhysRevD.1.1182}
L.~Susskind, ``Structure of hadrons implied by duality,'' {\em Phys. Rev. D},
  vol.~1, pp.~1182--1186, Feb 1970.

\bibitem{Nambu:1997wf}
Y.~Nambu, ``{Quark model and the factorization of the Veneziano amplitude},''
  in {\em {International Conference on Symmetries and Quark Models, Wayne State
  U., Detroit}}, pp.~269--278, 1997.

\bibitem{Green:1984sg}
M.~B. Green and J.~H. Schwarz, ``\textit{Anomaly Cancellation in Supersymmetric
  D=10 Gauge Theory and Superstring Theory},'' {\em Phys. Lett. B}, vol.~149,
  pp.~117--122, 1984.

\bibitem{gross1985heterotic}
D.~J. Gross, J.~A. Harvey, E.~Martinec, and R.~Rohm, ``Heterotic string,'' {\em
  Physical Review Letters}, vol.~54, no.~6, p.~502, 1985.

\bibitem{candelas1985vacuum}
P.~Candelas, G.~T. Horowitz, A.~Strominger, and E.~Witten, ``Vacuum
  configurations for superstrings,'' {\em Nuclear Physics B}, vol.~258,
  pp.~46--74, 1985.

\bibitem{th93}
G.~{'t Hooft}, ``{\textit{Dimensional Reduction in Quantum Gravity}},'' {\em
  arXiv e-prints}, pp.~gr--qc/9310026, Oct. 1993.

\bibitem{th94}
C.~R. Stephens, G.~Hooft, and B.~F. Whiting, ``{\textit{Black hole evaporation
  without information loss}},'' {\em Classical and Quantum Gravity}, vol.~11,
  no.~3, p.~621, 1994.

\bibitem{witten1995}
E.~Witten, ``String theory dynamics in various dimensions,'' 1995.

\bibitem{string1}
J.~Polchinski, {\em String theory}, vol.~Volume 1.
\newblock CUP, 1998.

\bibitem{susskind1994}
L.~Susskind, ``{\textit{The World as a Hologram}},'' {\em arXiv preprint
  hep-th/9409089}, 1994.

\bibitem{maldacena1997}
J.~M. Maldacena, ``{\textit{The Large N Limit of Superconformal Field Theories
  and Supergravity}},'' 1997.

\bibitem{aharony1999large}
O.~Aharony, S.~S. Gubser, J.~Maldacena, H.~Ooguri, and Y.~Oz, ``Large n field
  theories, string theory and gravity,'' 1999.

\bibitem{bfss}
T.~Banks, W.~Fischler, S.~H. Shenker, and L.~Susskind, ``M theory as a matrix
  model: A conjecture,'' 1996.

\bibitem{string3}
M.~Hanada, ``Monte carlo approach to the string/m-theory,'' 2012.

\bibitem{string4}
C.-H. Liu, ``Azumaya noncommutative geometry and d-branes - an origin of the
  master nature of d-branes,'' 2011.

\bibitem{Mmm}
E.~Dreyer, ``String theory at first glance,'' tech. rep., Simon Fraser
  University, Canada, 04 2017.

\bibitem{M-theory2}
R.~J. Szabo, {\em An Introduction to String Theory and D-brane Dynamics: With
  Problems and Solutions, Second Edition}.
\newblock Imperial College Press, 2~ed., 2011.

\bibitem{duality}
S.~Forste and J.~Louis, ``Duality in string theory,'' 1996.

\bibitem{string2}
S.~J. Becker~K., Becker~M., {\em String Theory and M-Theory}.
\newblock CUP, 2007.

\bibitem{SUGRA}
A.~Adeifeoba, ``Brane solutions in supergravity and the near-horizon
  geometries,'' Sep 2018.

\bibitem{anagnostopoulos2007}
K.~N. Anagnostopoulos, M.~Hanada, J.~Nishimura, and S.~Takeuchi, ``Monte carlo
  studies of supersymmetric matrix quantum mechanics with sixteen supercharges
  at finite temperature,'' 2007.

\bibitem{luminet2016holographic}
J.-P. Luminet, ``The holographic universe,'' 2016.

\bibitem{bousso2002holographic}
R.~Bousso, ``The holographic principle,'' 2002.

\bibitem{ads-cft}
I.~Kirsch, ``Generalizations of the ads/cft correspondence,'' 2004.

\bibitem{conden}
Y.-W.~S. Jan~Zaanen, Yan~Liu, {\em {\textit{Holographic Duality in Condensed
  Matter Physics}}}.
\newblock Cambridge University Press, 2016.

\bibitem{hartnoll}
S.~A. Hartnoll, A.~Lucas, and S.~Sachdev, {\em {\textit{Holographic quantum
  matter}}}.
\newblock MIT press, 2018.

\bibitem{ammon}
J.~E. Martin~Ammon, {\em {\textit{Gauge/Gravity Duality: Foundations and
  Applications}}}.
\newblock Cambridge University Press, 2015.

\bibitem{adler}
A.~Adler, {\em {\textit{The Individual Psychology of Alfred Adler: A Systematic
  Presentation in Selections from His Writings}}}.
\newblock Harper Perennial, 1964, Dec (p.259).

\bibitem{makeenkothree}
Y.~{Makeenko}, ``{\emph{Three Introductory Lectures in Helsinki on Matrix
  Models of Superstrings}},'' {\em arXiv e-prints}, pp.~hep--th/9704075, Apr.
  1997.

\bibitem{tanwar}
N.~Tanwar, ``Monte carlo simulations of bfss and ikkt matrix models,'' 2020.

\bibitem{catterall2008}
S.~Catterall and T.~Wiseman, ``Black hole thermodynamics from simulations of
  lattice yang-mills theory,'' 2008.

\bibitem{kawahara2007}
N.~Kawahara, J.~Nishimura, and S.~Takeuchi, ``High temperature expansion in
  supersymmetric matrix quantum mechanics,'' 2007.

\bibitem{kadoh2015}
D.~Kadoh and S.~Kamata, ``Gauge/gravity duality and lattice simulations of one
  dimensional sym with sixteen supercharges,'' 2015.

\bibitem{filev2015}
V.~G. Filev and D.~O'Connor, ``The bfss model on the lattice,'' 2015.

\bibitem{hanada-test}
E.~Berkowitz, E.~Rinaldi, M.~Hanada, G.~Ishiki, S.~Shimasaki, and P.~Vranas,
  ``Precision lattice test of the gauge/gravity duality at large-$n$,'' 2016.

\bibitem{hanada2008higher}
M.~Hanada, Y.~Hyakutake, J.~Nishimura, and S.~Takeuchi, ``Higher derivative
  corrections to black hole thermodynamics from supersymmetric matrix quantum
  mechanics,'' 2008.

\bibitem{hawking1975}
S.~W. Hawking, ``Particle creation by black holes,'' {\em Comm. Math. Phys.},
  vol.~43, no.~3, pp.~199--220, 1975.

\bibitem{small1}
M.~Hanada, Y.~Hyakutake, G.~Ishiki, and J.~Nishimura, ``Numerical tests of the
  gauge/gravity duality conjecture for d0-branes at finite temperature and
  finite n,'' 2016.

\bibitem{small2}
M.~Hanada, Y.~Hyakutake, G.~Ishiki, and J.~Nishimura, ``Holographic description
  of quantum black hole on a computer,'' 2013.

\bibitem{thooft2002large}
G.~'t~Hooft, ``Large n,'' 2002.

\bibitem{coleman1985}
S.~Coleman, {\em 1/N}, p.~351–402.
\newblock Cambridge University Press, 1985.

\bibitem{manohar1998large}
A.~V. Manohar, ``Large n qcd,'' 1998.

\bibitem{witten1979large}
E.~Witten, ``{Baryons in the 1/n Expansion},'' {\em Nucl. Phys. B}, vol.~160,
  pp.~57--115, 1979.

\bibitem{Drouffe:1979dh}
J.~Drouffe, G.~Parisi, and N.~Sourlas, ``{Strong Coupling Phase in Lattice
  Gauge Theories at Large Dimension},'' {\em Nucl. Phys. B}, vol.~161,
  pp.~397--416, 1979.

\bibitem{DROUFFE19831}
J.-M. Drouffe and J.-B. Zuber, ``Strong coupling and mean field methods in
  lattice gauge theories,'' {\em Physics Reports}, vol.~102, no.~1, pp.~1 --
  119, 1983.

\bibitem{kabat2000black}
D.~Kabat, G.~Lifschytz, and D.~A. Lowe, ``Black hole thermodynamics from
  calculations in strongly-coupled gauge theory,'' 2000.

\bibitem{kabat2001black}
D.~Kabat, G.~Lifschytz, and D.~A. Lowe, ``Black hole entropy from
  non-perturbative gauge theory,'' 2001.

\bibitem{gauss1}
G.~Mandal, M.~Mahato, and T.~Morita, ``Phases of one dimensional large n gauge
  theory in a 1/d expansion,'' 2009.

\bibitem{grav0}
J.~Maldacena and A.~Milekhin, ``To gauge or not to gauge?,'' 2018.

\bibitem{grav1}
J.~Maldacena, M.~M. Sheikh-Jabbari, and M.~V. Raamsdonk, ``Transverse
  fivebranes in matrix theory,'' 2002.

\bibitem{grav2}
N.~Itzhaki, J.~M. Maldacena, J.~Sonnenschein, and S.~Yankielowicz,
  ``Supergravity and the large n limit of theories with sixteen supercharges,''
  1998.

\bibitem{hyakutake2013quantum}
Y.~Hyakutake, ``Quantum near horizon geometry of black 0-brane,'' 2013.

\bibitem{ydricomputational}
B.~Ydri, ``Computational physics: an introduction to monte carlo simulations of
  matrix field theory,'' {\em arXiv preprint arXiv:1506.02567}, 2015.

\bibitem{hanadamarkov}
M.~Hanada, ``Markov chain monte carlo for dummies,'' {\em arXiv preprint
  arXiv:1808.08490}, 2018.

\bibitem{anosh}
A.~Joseph, {\em Markov Chain Monte Carlo Methods in Quantum Field Theories: A
  Modern Primer (SpringerBriefs in Physics)}.
\newblock SpringerBriefs in Physics, Springer, 1st ed. 2020~ed., 2020.

\bibitem{ydriblog}
B.~{Ydri}, ``{\emph{Lattice QFT (of Matrix Models)}}.'' \url
  {http://badisydri.blogspot.com/2019/10/lattice-qft-m-atrix-theory.html}.
\newblock October 02, 2019.

\bibitem{kawahara2007phase}
N.~Kawahara, J.~Nishimura, and S.~Takeuchi, ``Phase structure of matrix quantum
  mechanics at finite temperature,'' 2007.

\bibitem{ydri2020}
B.~Ydri, ``Two approaches to quantum gravity and m-(atrix) theory at large
  number of dimensions,'' 2020.

\bibitem{ydri2017review}
B.~Ydri, ``Review of m(atrix)-theory, type iib matrix model and matrix string
  theory,'' 2017.

\bibitem{hanada2019partial}
M.~Hanada, G.~Ishiki, and H.~Watanabe, ``Partial deconfinement in gauge
  theories,'' 2019.

\bibitem{myers-perry}
R.~C. {Myers} and M.~J. {Perry}, ``{Black holes in higher dimensional
  space-times},'' {\em Annals of Physics}, vol.~172, pp.~304--347, Dec. 1986.

\bibitem{kaluza}
T.~{Kaluza}, ``{Zum Unit{\"a}tsproblem der Physik},'' {\em Sitzungsberichte der
  K{\"o}niglich Preu{\ss}ischen Akademie der Wissenschaften (Berlin},
  pp.~966--972, Jan. 1921.

\bibitem{klein}
O.~Klein, ``The atomicity of electricity as a quantum theory law,'' {\em
  Nature}, vol.~118, no.~2971, pp.~516--516, 1926.

\bibitem{duff1996black}
M.~J. Duff, H.~Lu, and C.~N. Pope, ``The black branes of m-theory,'' 1996.

\bibitem{mohaupt2000black}
T.~Mohaupt, ``Black holes in supergravity and string theory,'' 2000.

\bibitem{galtsov2005general}
D.~Gal'tsov, S.~Klevtsov, D.~Orlov, and G.~Clement, ``More on general $p$-brane
  solutions,'' 2005.

\bibitem{Tangherlini:1963bw}
F.~Tangherlini, ``{Schwarzschild field in n dimensions and the dimensionality
  of space problem},'' {\em Nuovo Cim.}, vol.~27, pp.~636--651, 1963.

\bibitem{lemos1994twodimensional}
J.~P.~S. Lemos, ``Two-dimensional black holes and planar general relativity,''
  1994.

\bibitem{lemos1994cylindrical}
J.~P.~S. Lemos, ``Cylindrical black hole in general relativity,'' 1994.

\bibitem{emparan2008black}
R.~Emparan and H.~S. Reall, ``Black holes in higher dimensions,'' 2008.

\bibitem{horo-stro}
G.~T. {Horowitz} and A.~{Strominger}, ``{Black strings and p-branes},'' {\em
  Nuclear Physics B}, vol.~360, pp.~197--209, Aug. 1991.

\bibitem{gregory1993black}
R.~Gregory and R.~Laflamme, ``Black strings and p-branes are unstable,'' 1993.

\bibitem{unstablestring2003}
M.~Choptuik, L.~Lehner, I.~I.~n. Olabarrieta, R.~Petryk, F.~Pretorius, and
  H.~Villegas, ``Towards the final fate of an unstable black string,'' {\em
  Phys. Rev. D}, vol.~68, p.~044001, Aug 2003.

\bibitem{PhysRevLett.87.135001}
R.~A. Moyer, G.~R. Tynan, C.~Holland, and M.~J. Burin, ``Increased nonlinear
  coupling between turbulence and low-frequency fluctuations at the
  $l\ensuremath{-}h$ transition,'' {\em Phys. Rev. Lett.}, vol.~87, p.~135001,
  Sep 2001.

\bibitem{gubser2001nonuniform}
S.~S. Gubser, ``On non-uniform black branes,'' 2001.

\bibitem{sorkin2004critical}
E.~Sorkin, ``A critical dimension in the black-string phase transition,'' 2004.

\bibitem{sorkin2006nonuniform}
E.~Sorkin, ``Nonuniform black strings in various dimensions,'' 2006.

\bibitem{kol2002topology}
B.~Kol, ``Topology change in general relativity and the black-hole black-string
  transition,'' 2002.

\bibitem{wiseman2002black}
T.~Wiseman, ``From black strings to black holes,'' 2002.

\bibitem{kol2003evidence}
B.~Kol and T.~Wiseman, ``Evidence that highly non-uniform black strings have a
  conical waist,'' 2003.

\bibitem{kleihaus2007interior}
B.~Kleihaus and J.~Kunz, ``Interior of nonuniform black strings,'' 2007.

\bibitem{susskind1998}
L.~Susskind, ``Matrix theory black holes and the gross witten transition,''
  1998.

\bibitem{aharony2004black}
O.~Aharony, J.~Marsano, S.~Minwalla, and T.~Wiseman, ``Black hole-black string
  phase transitions in thermal 1+1-dimensional supersymmetric yang-mills theory
  on a circle,'' 2004.

\bibitem{harmark2004new}
T.~Harmark and N.~A. Obers, ``New phases of near-extremal branes on a circle,''
  2004.

\bibitem{obers2008black}
N.~A. Obers, ``Black holes in higher-dimensional gravity,'' 2008.

\bibitem{kudoh2004connecting}
H.~Kudoh and T.~Wiseman, ``Connecting black holes and black strings,'' 2004.

\bibitem{hagedorn1}
T.~G. Mertens, H.~Verschelde, and V.~I. Zakharov, ``Hagedorn temperature and
  physics of black holes,'' 2016.

\bibitem{aharony2003hagedorn}
O.~Aharony, J.~Marsano, S.~Minwalla, K.~Papadodimas, and M.~V. Raamsdonk, ``The
  hagedorn/deconfinement phase transition in weakly coupled large n gauge
  theories,'' 2003.

\bibitem{eguchi-kawai}
T.~Eguchi and H.~Kawai, ``Reduction of dynamical degrees of freedom in the
  large-$n$ gauge theory,'' {\em Phys. Rev. Lett.}, vol.~48, pp.~1063--1066,
  Apr 1982.

\bibitem{kawahara2007fuzzy}
N.~Kawahara, J.~Nishimura, and S.~Takeuchi, ``Exact fuzzy sphere thermodynamics
  in matrix quantum mechanics,'' 2007.

\bibitem{madore1992fuzzy}
J.~Madore, ``The fuzzy sphere,'' {\em Classical and Quantum Gravity}, vol.~9,
  no.~1, p.~69, 1992.

\bibitem{maceda2011fuzzy}
M.~Maceda, ``Fuzzy physics: A brief overview of noncommutative geometry in
  physics,'' in {\em AIP Conference Proceedings}, vol.~1396, pp.~65--74,
  American Institute of Physics, 2011.

\bibitem{balachandran2005lectures}
A.~Balachandran, S.~Kurkcuoglu, and S.~Vaidya, ``Lectures on fuzzy and fuzzy
  susy physics,'' {\em arXiv preprint hep-th/0511114}, 2005.

\bibitem{balachandran2006noncommutative}
A.~P. Balachandran, B.~A. Qureshi, {\em et~al.}, ``Noncommutative geometry:
  Fuzzy spaces, the groenewold-moyal plane,'' {\em SIGMA. Symmetry,
  Integrability and Geometry: Methods and Applications}, vol.~2, p.~094, 2006.

\bibitem{berenstein2003strings}
D.~Berenstein, J.~Maldacena, and H.~Nastase, ``Strings in flat space and pp
  waves from \&caln;= 4 super yang-mills,'' {\em -}, 2003.

\bibitem{KOWALSKIGLIKMAN1984194}
J.~Kowalski-Glikman, ``Vacuum states in supersymmetric kaluza-klein theory,''
  {\em Physics Letters B}, vol.~134, no.~3, pp.~194 -- 196, 1984.

\bibitem{li1999short}
M.~Li and T.~Yoneya, ``Short-distance space-time structure and black holes in
  string theory: A,'' in {\em Short Review of the Present Status,
  hep-th/9806240, Jour. Chaos, Solitons and Fractals}, Citeseer, 1999.

\bibitem{jevicki1999non}
A.~Jevicki and S.~Ramgoolam, ``Non commutative gravity from the ads/cft
  correspondence,'' {\em Journal of High Energy Physics}, vol.~1999, no.~04,
  p.~032, 1999.

\bibitem{dubois1989gauge}
M.~Dubois-Violette, J.~Madore, and R.~Kerner, ``Gauge bosons in a
  noncommutative geometry,'' {\em Physics Letters B}, vol.~217, no.~4,
  pp.~485--488, 1989.

\bibitem{schomerus1999d}
V.~Schomerus, ``D-branes and deformation quantization,'' {\em Journal of High
  Energy Physics}, vol.~1999, no.~06, p.~030, 1999.

\bibitem{delgadilo}
R.~Delgadillo-Blando, D.~O'Connor, and B.~Ydri, ``{Matrix Models, Gauge Theory
  and Emergent Geometry},'' 2008.

\bibitem{hanada-code}
M.~{Hanada}, ``{\emph{Monte Carlo simulation code for Matrix Model of
  M-theory}}.'' \url {https://sites.google.com/site/hanadamasanori/home/mmmm}.
\newblock apr. 2020.

\bibitem{GWW1}
D.~J. Gross and E.~Witten, ``Possible third-order phase transition in the large
  $n$ lattice gauge theory,'' {\em Phys. Rev. D}, vol.~21, pp.~446--453, Jan
  1980.

\bibitem{GWW2}
S.~R. Wadia, ``{$N$ = Infinity Phase Transition in a Class of Exactly Soluble
  Model Lattice Gauge Theories},'' {\em Phys. Lett. B}, vol.~93, pp.~403--410,
  1980.

\end{thebibliography}
\bibliographystyle{ieeetr}

\end{document}